\documentclass[useAMS,usenatbib]{mn2e}
\usepackage{graphicx}
\usepackage{times}
\usepackage{amssymb} 
\usepackage{txfonts}
\usepackage{appendix}
\usepackage{natbib}
\usepackage{longtable, portland}
\usepackage{supertabular}
\usepackage{rotating}
\usepackage{array}
\usepackage{caption}
\usepackage{gensymb}

\title[The ALHAMBRA survey: morphological classification]{The ALHAMBRA survey: reliable morphological catalogue of 22,051 early- and late-type galaxies}
\author[Povi\'c et al.]{M. Povi\'c$^{1}$\thanks{E-mail:
mpovic@iaa.es}, M. Huertas-Company$^{2, 3}$,  J. A. L. Aguerri$^{4}$, I. M\'arquez$^{1}$, J. Masegosa$^{1}$, \newauthor C. Husillos$^{1}$, A. Molino$^{1}$, D. Crist\'obal-Hornillos$^{5}$, J. Perea$^{1}$, N. Ben\'itez$^{1}$, A. del Olmo$^{1}$, \newauthor Y. Jim\'enez-Teja$^{1}$, M. Moles$^{5, 1}$, E. Alfaro$^{1}$, T. Aparicio-Villegas$^{6, 1}$, B. Ascaso$^{1}$, \newauthor T. Broadhurst$^{7, 8}$, J. Cabrera-Ca\~no$^{9}$, F. J. Castander$^{10}$, J. Cepa$^{11, 4}$, M. Cervi\~no$^{1, 4}$, \newauthor M. Fern\'andez Lorenzo$^{1}$, A. Fern\'andez-Soto$^{12, 13}$, R. M. Gonz\'alez Delgado$^{1}$, L. Infante$^{14}$, \newauthor C. L\'opez-Sanjuan$^{5}$, V. J. Mart\'inez$^{15, 13}$, I. Matute$^{1}$, I. Oteo$^{4, 11}$, A. M. P\'erez-Garc\'ia$^{4, 11, 16}$, \newauthor F. Prada$^{1}$, and J. M. Quintana$^{1}$\\ \\
$^{1}$Instituto de Astrof\'isica de Andaluc\'ia (IAA-CSIC), Granada, Spain \\
$^{2}$GEPI, Paris-Meudon Observatory, Meudon, France\\ 
$^{3}$University of Paris, Paris, France\\ 
$^{4}$Instituto de Astrof\'isica de Canarias (IAC), La Laguna, Tenerife, Spain\\ 
$^{5}$Centro de Estudios de F\'isica del Cosmos de Arag\'on (CEFCA), Teruel, Spain\\
$^{6}$Observat\'orio Nacional-MCT, Rua José Cristino, 77. CEP 20921-400, Rio de Janeiro-RJ, Brazil\\
$^{7}$Department of Theoretical Physics, University of Basque Country, UPV/EHU, P.O. Box 644, E-48080 Bilbao, Spain\\
$^{8}$IKERBASQUE, Basque Foundation for Science, Alameda Urquijo 36-5, E-48008 Bilbao, Spain\\
$^{9}$Facultad de F\'isica. Departamento de F\'isica At\'omica, Molecular y Nuclear, Universidad de Sevilla, Sevilla, Spain\\
$^{10}$Institut de Ci\`encies de l'Espai, IEEC/CSIC, Barcelona, Spain\\ 
$^{11}$Departamento de Astrof\'isica, Facultad de F\'isica, Universidad de la Laguna, La Laguna, Spain \\ 
$^{12}$Instituto de F\'isica de Cantabria (CSIC-UC), Santander, Spain\\ 
$^{13}$Unidad Asociada Observatori Astron\`omic (IFCA - UV), Valencia, Spain\\
$^{14}$Departamento de Astronom\'ia, Pontificia Universidad Cat\'olica, Santiago de Chile, Chile\\ 
$^{15}$Observatori Astron\`omic de la Universitat de Val\`encia, Valencia, Spain\\ 
$^{16}$Asociac\'ion ASPID, Apartado de Correos 412, La Laguna, Tenerife, Spain\\
}
\begin{document}

\date{Accepted ??. Received ??; in original form ??}

\pagerange{\pageref{firstpage}--\pageref{lastpage}} \pubyear{2012}

\maketitle

\label{firstpage}

\begin{abstract}
ALHAMBRA is a photometric survey designed to trace the cosmic evolution and cosmic variance. It covers a large area of $\sim$\,4\,deg$^2$ in 8 fields, where 7 fields overlap with other surveys, allowing to have complementary data in other wavelengths. All observations were carried out in 20 continuous, medium band (30\,nm width) optical and 3 near-infrared ($JHK$) bands, providing the precise measurements of photometric redshifts. In addition, morphological classification of galaxies is crucial for any kind of galaxy formation and cosmic evolution studies, providing the information about star formation histories, their environment and interactions, internal perturbations, etc. We present a morphological classification of $>$\,40,000 galaxies in the ALHAMBRA survey. We associate to every galaxy a probability to be early-type using the automated Bayesian code galSVM. Despite of the spatial resolution of the ALHAMBRA images ($\sim$\,1\,arcsec), for 22,051 galaxies we obtained the contamination by other type of less than 10\%. Of those, 1,640 and 10,322 galaxies are classified as early- (down to redshifts $\sim$\,0.5) and late-type (down to redshifts $\sim$\,1.0), respectively, with magnitudes F613W\,$\le$\,22.0. In addition, for magnitude range 22.0\,$<$\,F613W\,$\le$\,23.0 we classified other 10,089 late-type galaxies with redshifts $\le$\,1.3. We show that the classified objects populate the expected regions in the colour-mass and colour-magnitude planes. The presented dataset is especially attractive given the homogeneous multi wavelength coverage available in the ALHAMBRA fields, and is intended to be used in a variety of scientific applications. The low-contamination catalogue ($<$\,10\%) is made publicly available with the present paper.
\end{abstract}


\section{Introduction}
\label{sec_intro}

\indent One of the first steps in any research work is to group objects with common properties (e.g. shape, weight, color,...). This taxonomy is a powerful tool in order to understand the physics behind the formation and evolution of the studied objects. Probably the most popular classification of galaxies is based on the shapes or morphologies \citep[started with][]{reynolds20,hubble26,hubble36}. This first order classification has survived over time since the different morphological classes of galaxies have also different physical properties and probably different evolutionary tracks. In general, galaxies can be divided into two main classes: early-types (hereafter ET) and late-types (hereafter LT). ETs include elliptical and lenticular galaxies, while LTs include spirals and irregular galaxies. ETs appear to be a family of objects showing old stellar populations, spheroidal-like dynamical properties and a small fraction of cold gas whereas LTs are more gas-rich objects, present younger stellar populations, and are mainly rotation supported.

\indent Visual inspection is the traditional way to classify galaxies. By definition, it is subjective and not reproducible, but for bright and extended objects there is a general good agreement between different observers. They are however time consuming. In the past, galaxy samples contained from dozens up to hundreds of galaxies while present galaxy surveys have up to millions of galaxies. This makes impossible to give detailed morphological classifications unless a large amount of classifiers are involved \citep[see the Galaxy Zoo project;][]{lintott08}.

\indent Over the past years, different automated methods of morphological classification of galaxies have been developed. Automated classifications resolve the two main problems raised above. They provide indeed reproducible information, and the errors can be fully understood by using extensive simulations  \citep[e.g.][]{trujillo01,simard02,simard11}. In addition, modern galaxy classification algorithms are able to classify large samples of galaxies in a reasonable amount time.

\indent We usually distinguish between three broad groups of automated galaxy classifications: parametric based on galaxy physical parameters, non-parametric, and parametric based on mathematical parameters. Parametric classifications use parametric models in order to reproduce some galaxy measurements. One of the most popular parametric methods based on galaxy physical parameters classifies galaxies according to some properties of their structural parameters obtained by fitting the surface brightness \citep[e.g.,][]{vaucouleurs48,sersic63,prieto97,prieto01,peng02,simard02,souza04,aguerri04,aguerri05,mendez08,peng10,simard11}. On the other side, non-parametric galaxy classifications are based on the measurements of a set of galaxy parameters that correlate with the morphological types. These methods have the advantage that they assume non-parametric models and can hence classify regular and irregular objects. Several galaxy parameters have been used to discriminate between different morphological types, i.e colours \citep[e.g.][]{strateva01}, spectral properties \citep[e.g.][]{humason31,morgan57,baldwin81,folkes96,sanchez10}, or light distribution \citep[e.g.][]{abraham94,abraham96,abraham03,conselice00,lotz04,scarlata07}. Current works are led to more complex fitting models, using orthonormal mathematical bases to decompose the galaxies, and then correlating the physical properties of the objects with the coefficients of the decomposition by means of a principal component analysis. This is the case of the shapelets \citep{kelly04,kelly05,andrae11a}, or future works with the sersiclets \citep{ngan09,andrae11b}, and the CHEFs \citep{jimenez12}. The main advantage of these latter methods with regard to the former is their flexibility and reliability to reproduce every feature in the galaxies, and then to efficiently model every kind of morphology, including irregular objects. Moreover, they do not need any a priori knowledge and they do not impose any profile to fit the galaxies. However, these mathematical models are computationally more expensive and the parameters of the final decomposition do not provide any physical information. ALH data will be used for further development of the CHEFs method (Jim\'enez et al., in preparation). 

\indent  galSVM \citep{huertas08} provides a generalization of the non-parametric classifications by using an unlimited number of dimensions and providing a probabilistic output following a Bayesian approach (see also \citealp{fasano12} for a similar approach). The algorithm has been shown to be specially useful when dealing with low-resolution and high-redshift data\citep{huertas09} and has been successfully applied to several large samples at low and high redshift \citep{huertas09,huertas11}, including the Cosmological Evolution Survey \citep[COSMOS\footnote{http://cosmos.astro.caltech.edu/};][]{scoville07} and Sloan Digital Sky Survey \citep[SDSS\footnote{http://www.sdss.org/};][]{castander98} samples. \\
\indent In this paper we present the morphological classification of a large sample of galaxies from the Advanced Large Homogeneous Area Medium Band Redshift Astronomical, ALHAMBRA survey \citep[hereafter ALH; see][]{moles08}, located at different redshifts. Taking into account the resolution of ALH data ($\sim$\,1\,arcsec) and all the advantages of galSVM code mentioned above, we used this method for our morphological classification. ALH survey imaged $\sim$\,4\,deg$^2$ of the sky through 23 optical and near-infrared (NIR) filters. The large number of filters provide accurate photometric redshifts by fitting the Spectral Energy Distributions (hereafter SED) of the galaxies for about one million sources. This galaxy sample is ideal in order to study evolutionary properties of galaxies in the last 8 Gyr \citep{moles08,cristobal09,matute12,oteo13a,oteo13b}.

\indent The ALHAMBRA survey, data, and sample selection are introduced in Sec.~\ref{sec_observations}. The methodology used for morphological classification is described in Sec.~\ref{sec_morph_class}, and the main results are discussed in Sec.~\ref{sec_class_reliability}, \ref{sec_final_class}, and \ref{sec_prop}. Finally, Appendix~\ref{appendixA_cat} describes the content of the first published morphological catalogue of galaxies in the ALHAMBRA survey. 

\indent We assumed the following cosmological parameters throughout the paper: $\Omega_{\Lambda}$\,=\,0.7, $\Omega_{M}$\,=\,0.3, and H$_0$\,=\,70 km s$^{-1}$ Mpc$^{-1}$. Unless otherwise specified, all magnitudes are given in the AB system \citep{oke83}.

\section[]{Data}
\label{sec_observations}

\begin{figure*}
\centering
\begin{minipage}[c]{.49\textwidth}
\includegraphics[width=8.4cm,angle=0]{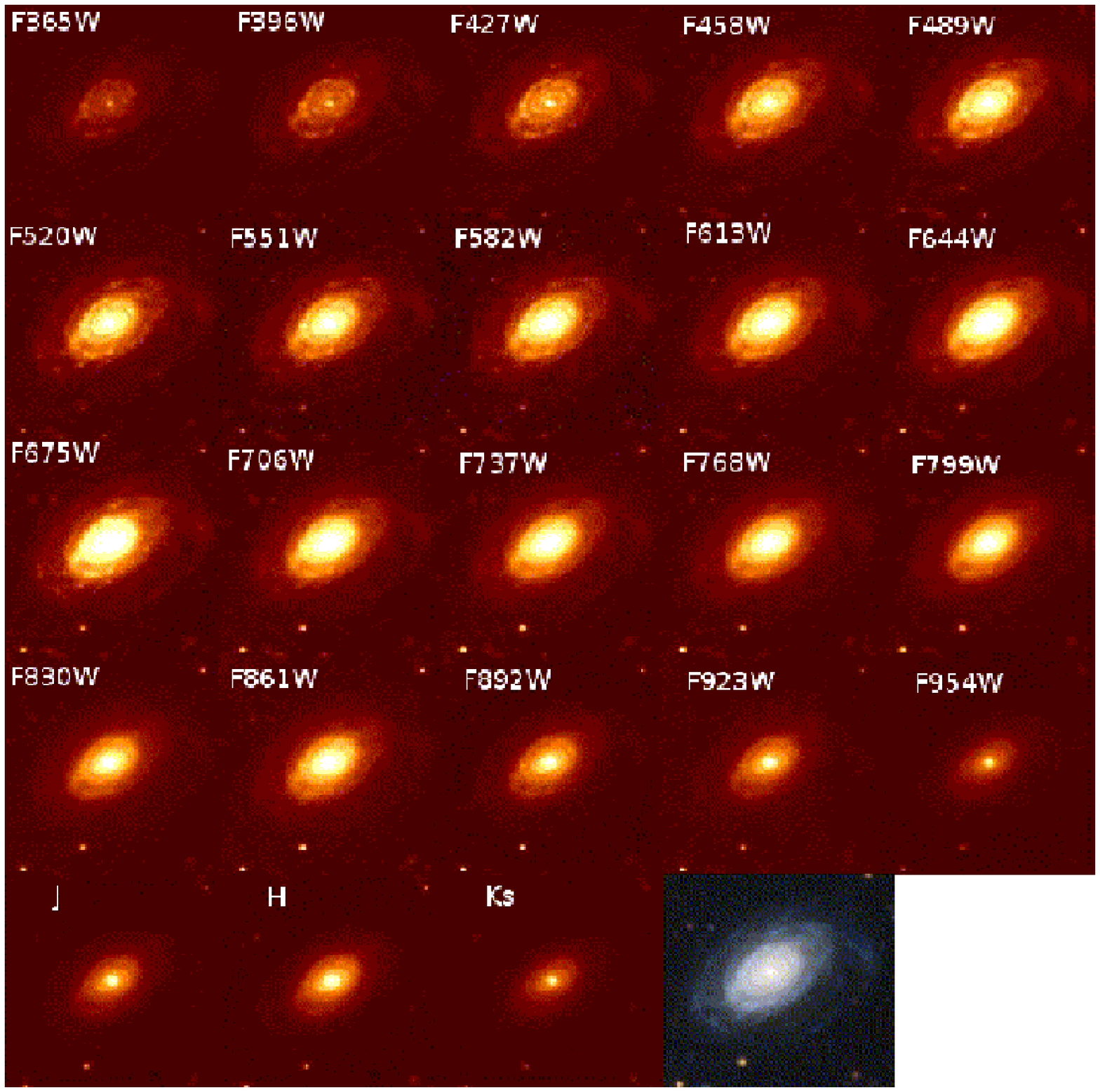}
\end{minipage}
\begin{minipage}[c]{.49\textwidth}
\includegraphics[width=8.35cm,angle=0]{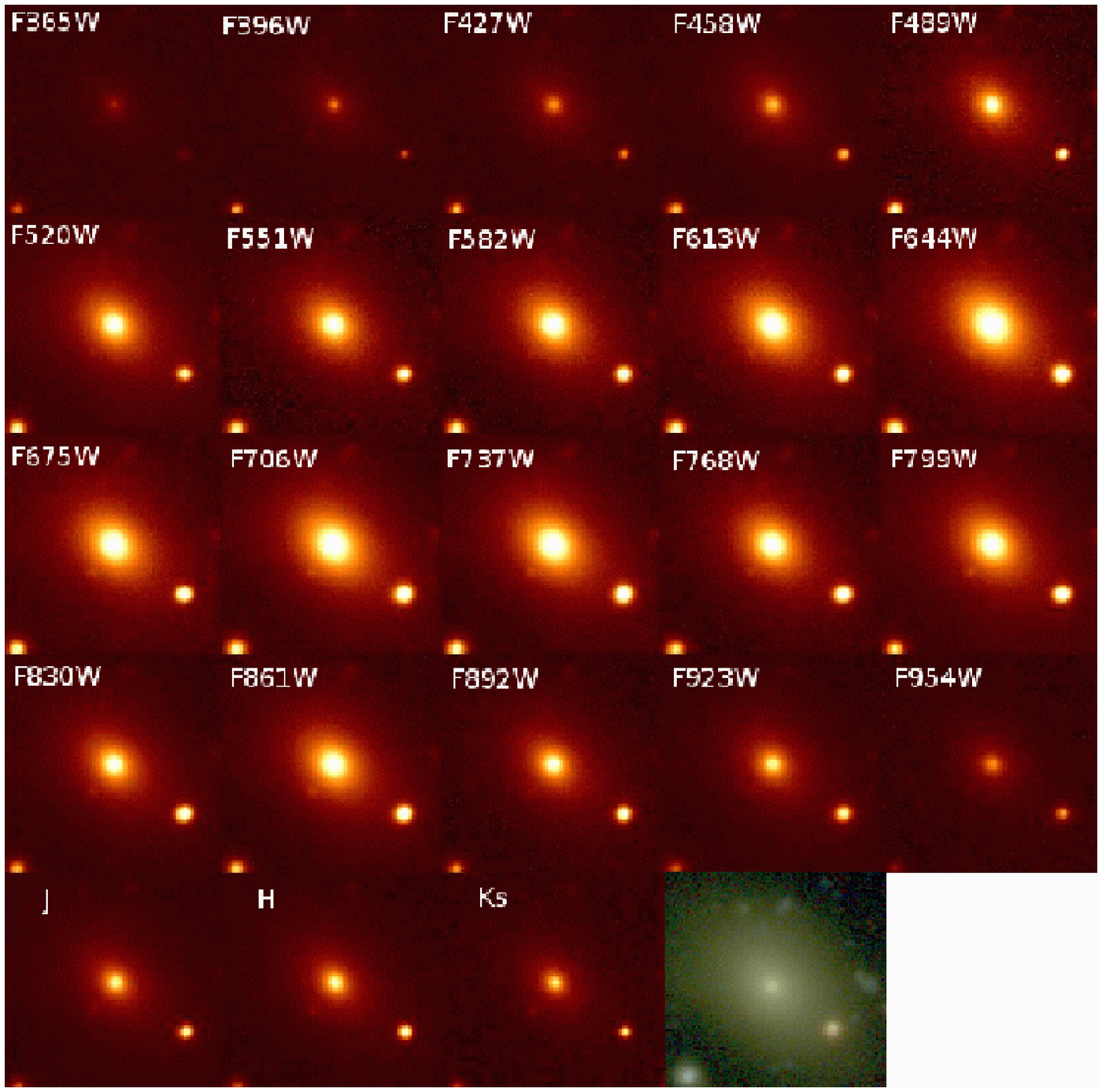}
\end{minipage}
\caption[ ]{Example of two galaxies observed with 20 optical and 3 near-infrared ALH filters.}
\label{fig_alh_example}
\end{figure*}

\indent The ALHAMBRA observations were carried out at the Calar Alto German-Spanish Astronomical Center (CAHA\footnote{http://www.caha.es/}), under the Spanish guaranteed time of 110 nights. Eight fields were observed in the northern hemisphere sky, having a seeing lower than 1.6\,arcsec (ranging mainly between 0.8 and 1.2\,arcsec). All fields (except ALH-1) have the multiwavelength information available from other (deep) extragalactic surveys (see \cite{moles08} and Table~\ref{tab_alhambra_obs}). Each ALH detection was observed in 23 bands, with 20 optical and three standard $JHK$ NIR filters. Optical range is covered in a continuous way from 340\,nm to 970\,nm, with non-overlapping and equal 30\,nm width medium bands \citep{aparicio10}. This set of optical filters was specially designed for the ALH survey to achieve a good accuracy of photometric redshifts of $\sim$\,$\delta$z\,/\,(1\,+\,z)=\,0.015 for galaxies brighter than F814W\,$\le$\,24.5, three times better than the one achieved when using the 4\,-\,5 filter systems \citep[for more information see][]{benitez09}. Figure~\ref{fig_alh_example} shows the example of two galaxies observed in all ALH bands. 

\begin{table*}
\begin{center}
\caption{ALHAMBRA observations used in this paper: ALHAMBRA fields and their central coordinates, number of objects and covered areas in the F613W band down to magnitudes 23.0, range of seeing of individual observations used to create the final ones, and the averaged seeing. 
\label{tab_alhambra_obs}}
\begin{tabular}{c | c | c | c | c | c | c }
\hline
\textbf{Field}&\textbf{RA (J2000)}&\textbf{DEC (J2000)}&\textbf{Num. of obj. in F613W band}&\textbf{area}&\textbf{min\,-\,max seeing}&\textbf{averaged seeing}\\ 
&(h m s)&($\degree$ $'$ $''$)&(at mag\,$\le$\,23.0)&(deg$^2$)&(arcsec)&(arcsec)\\
\hline
ALH-2/DEEP2&02 28 32.0& +00 47 00&14322&0.5&0.86\,-\,1.40&1.04\\ 
ALH-3/SDSS&09 16 20.0& +46 02 20&12508&0.5&0.70\,-\,1.18&0.89\\ 
ALH-4/COSMOS&10 00 28.6& +02 12 21&7104&0.25&1.06\,-\,1.32&1.17\\ 
ALH-5/HDF-N&12 35 00.0& +61 57 00&6274&0.25&0.95\,-\,1.40&1.23\\ 
ALH-6/GROTH&14 16 38.0& +52 25 05&13614&0.5&0.81\,-\,1.30&1.11\\ 
ALH-7/ELAIS-N1&16 12 10.0& +54 30 00&15887&0.5&0.84\,-\,1.40&1.04\\ 
ALH-8/SDSS&23 45 50.0& +15 34 50&14128&0.5&0.72\,-\,1.40&0.91\\ 
\hline
\end{tabular}
\end{center}
\end{table*} 

\indent Optical data were observed with Large Area Imager for Calar Alto (LAICA\footnote{http://www.caha.es/CAHA/Instruments/LAICA/index.html}), with the total exposure time of 100000\,sec per pointing. On the other hand, OMEGA2000\footnote{http://w3.caha.es/CAHA/Instruments/O2000/index2.html} instrument was used in the NIR, with the total exposure time of 60000\,sec per pointing. Data reduction was carried out using the standard set of IRAF\footnote{IRAF is distributed by the National Optical Astronomy Observatory (NOAO), which is operated by the Association of Universities for Research in Astronomy (AURA) under cooperative agreement with the National Science Foundation (NSF).} packages. Table~\ref{tab_alhambra_obs} gives the summary of ALH observations used in this work.

\indent Source detection was performed by means of the SExtractor (v. 2.8.6) code \citep{bertin96}. Both, detection of sources and creation of photometric catalogues, were carried out in all 23 bands, where each catalogue contains more than 180 photometric, astrometric and basic morphological parameters. More than 670,000 sources were detected in all ALH survey, with a photometric completeness of $r$\,$\sim$\,25.0 (corresponds to SDSS $r$ band constructed from five individual ALH bands). All additional information respect to the source detection and creation of photometric catalogues can be found in \cite{husillos13}, while the catalogues are publicly available via the survey webpage\footnote{http://alhambrasurvey.com/}. 

\indent For photometric redshift estimations (hereafter photo-z), the Bayesian Photometric Redshift code was used \citep[BPZ;][]{benitez00,benitez04}. BPZ was run on a separated point spread function (PSF) corrected photometry and a new library of templates \citep{benitez13} was implemented, composed by eleven SEDs (4 elliptical, 1 lenticular, 1 Sbc, 1 Scd, and 4 starburst galaxies), originally drawn from PEGASE \citep{fioc97}, but then optimized using the FIREWORKS photometry and spectroscopic redshifts from \cite{wuyts08}. Based on a sample of $\sim$\,7000 galaxies with known spectroscopic redshifts, drawn from different surveys (see Table~\ref{tab_alhambra_obs}), the final performance of resulting photo-zs was evaluated. The expected accuracy for galaxies with magnitudes\,$<$\,23.0 in the constructed F814W\,nm band is $\sim$\,$\delta$z\,/\,(1\,+\,z)=\,0.011 with a small fraction of catastrophic outliers no higher than 3\%. In this work we used the BPZ measured photo-zs, and the same photometry that was used in photo-z estimations. All information about the photo-z measurements and catalogues can be found in \cite{molino13}.

\subsection{Sample selection}
\label{subsec_real_sample}

\indent We carried out the morphological classification on a well defined sample, free of stars, having reliable photo-z and photometric measurements. In this Section we describe the different selection criteria applied to obtain the final working sample.
\begin{itemize}
\item \textit{Extended\,-\,point like source separation.} To discriminate between the extended and point-like sources we used the classification provided by \cite{husillos13}. In all observations, a sample of more than 2,000 real, bright, unsaturated, and geometrically circular point-like ALH sources was used, scaled to lower random fluxes, and injected into the empty regions of optical image. After this, the sources were recovered, using the same parametrisation as applied in source detection. Point-like sources are considered as successfully recovered when detected at distances lower than 1\,arcsec from previous injection coordinates. The same procedure was implemented for extended sources. Finally, the locus of point-like and extended sources was defined in the apparent magnitude and surface brightness space (MAG\_AUTO\footnote{Kron-like elliptical aperture magnitude (see Bertin's and Holwerda's SExtractor manuals at https://www.astromatic.net and http://mensa.ast.uct.ac.za/$\sim$holwerda/Site/Source\_Extractor.html, respectively).} and MU\_MAX\footnote{Peak surface brightness above background (see above manuals).} parameters, respectively, obtained by SExtractor). By plotting each locus, it is possible to estimate the point-like source contamination in a quantitative way. GEOM\_CLASS\_STAR parameter was defined in this way, having value equal to 1 for point-like, and equal to 0 for extended sources. Unclassified sources have GEOM\_CLASS\_STAR\,=\,99. Taking into account the ALH resolution, it was confirmed that this method works well down to magnitudes $r$\,=\,23.0, which is in a good agreement with our magnitude selection criteria (see below).\\ 
\indent To select the extended sources, we used the GEOM\_CLASS\_STAR\,=\,0. With this criteria, using the comparisons with the Hubble Space Telescope (HST) data from the COSMOS survey, we estimated to have in the selected sample contamination of point-like sources lower than 1\% down to magnitudes 22.0 in the F613W band, and of 5\,-\,7\% between magnitudes 22.0 and 23.0.
\item \textit{Photo-z selection.} We used two criteria to select the sources with good photo-z measurements. First, we only selected those sources detected in all ALH filters. And second, we selected sources with BPZ\_ODDS\footnote{Defines the redshift confidence limits, where galaxies
with higher ODDS have a more secure redshift estimations. \citep[see][for more information]{benitez00}.} parameter above 0.2. With these two criteria, we expect to have less than 3\% of outliers \citep{molino13}. We checked that requiring the objects to be detected in all ALH bands introduces only a small selecion bias affecting the sample of red galaxies at z\,$>$\,0.4 and magnitudes $<$\,22 in the F613W band, a range in which our method is not able to efficiently select early-types (see Sec.~\ref{sec_class_reliability}).  
\item \textit{Magnitude selection.} We selected only objects with magnitudes $\le$\,23.0 in the F613W filter, and with magnitude errors $<$\,0.5. Above this magnitude limit, the reliability of signal-to-noise (S/N) measurements, photo-z estimations, and  geometrical extended/point-source classifications decrease significantly. Selection of the F613W filter is explained in Sec.~\ref{subsec_galsvm_config_classifiaction}. In comparison with previous criteria, magnitude selection is the most restrictive one. 
\item \textit{Flag tests.} In the sample selected through previous conditions, 57\% of sources are 'good detections' (SExtractor FLAG parameter =\,0), 27\% are possibly blended sources (FLAG\,=\,2), 15\%, plus blending, have close neighbours or bad pixels (FLAG\,=\,3), while $<$\,1\% have other FLAG values. We classified morphologically all galaxies independently of their FLAG parameter, including close/interacting systems. However, the final statistics (Sec.~\ref{sec_final_class} and \ref{sec_prop}), and the published catalogue (Appendix~\ref{appendixA_cat}) include only sources with FLAG values 0 and 2 (see Sec.~\ref{sec_class_reliability}).
\end{itemize}
\indent The final selected sample to be morphologicaly classified has 43,665 galaxies.

\section[]{Morphological classification of galaxies}
\label{sec_morph_class}

\subsection{General methodology}
\label{subsec_method}

\begin{figure*}
\centering
\begin{minipage}[c]{.49\textwidth}
\includegraphics[width=9.0cm,angle=0]{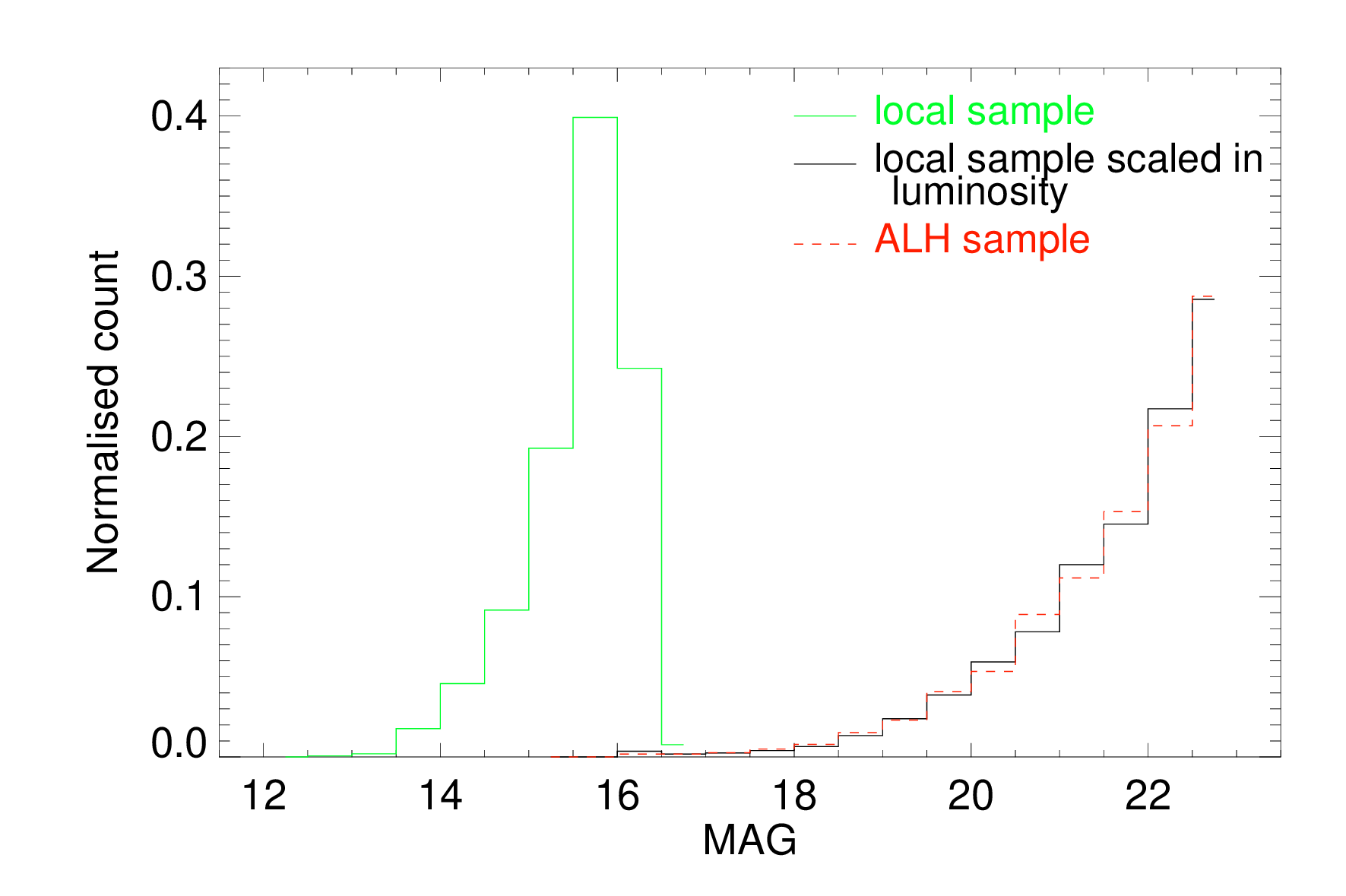}
\end{minipage}
\begin{minipage}[c]{.49\textwidth}
\includegraphics[width=9.0cm,angle=0]{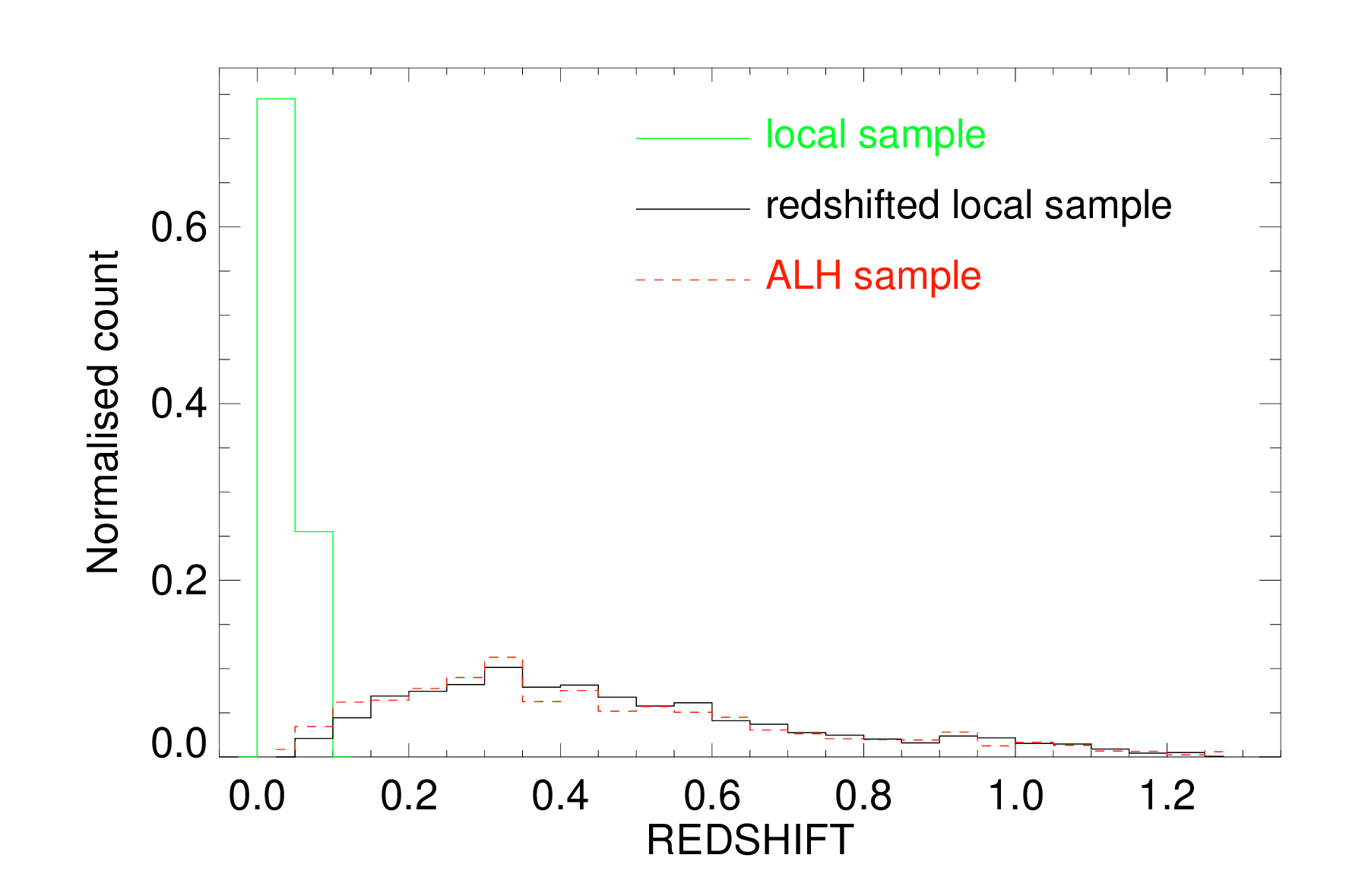}
\end{minipage}
\caption[ ]{Magnitude and redshift normalised distributions of local sample used by galSVM (green solid lines). This sample was scaled in luminosity and redshifted (black solid lines) to match the distributions of ALH sample (red dashed lines). Showed example corresponds to distributions in the ALH-7 field.}
\label{fig_histos_mag_z_local_alh}
\end{figure*}

\indent The main tool used in this work to estimate morphologies is galSVM, a non-parametric support vector machine (SVM) based code \citep{huertas08,huertas09,huertas11}. Basically, galSVM uses a training set of local galaxies with known visual morphologies to train the SVM that is then applied to the dataset to be classified (see Sec.~\ref{subsec_local_sample}). Galaxies from the training sample are redshifted and scaled in luminosity to match the magnitude counts and redshift distribution of the ALH sample, resampled with the ALH pixel scale ($\sim$\,0.222\,arcsec/pix), and finally dropped in a real ALH background. Figure~\ref{fig_histos_mag_z_local_alh} shows an example of redshift and magnitude distributions of local sample before and after being redshifted and scaled in luminosity (green and black solid lines), and corresponding distributions of ALH-7 sample (red dashed lines).\\
\indent We measured 7 morphological parameters on this simulated dataset (and on the ALH sample later on), and use them simultaneously to train the vector machine: 
\begin{itemize}
\item Ellipticity
\item Abraham concentration index (CABR) - ratio between the fluxes at 30\% and 90\% of the radius \citep{abraham96}
\item Conselice-Bershady concentration index (CCON) - ratio between circular radii containing 20\% and 80\% of the total flux \citep{bershady00}
\item Gini (GINI) - cumulative distribution function of galaxy's pixel values \citep{abraham03} 
\item Asymmetry (ASYM) - measures the degree of asymmetry in the light distribution \citep{conselice00}
\item Smoothnes (SMOOTH)- measures the relevance of small-scale structures \citep{conselice00}, and 
\item M$_{20}$ moment of light (M20) - flux in each pixel multiplied by the squared distance to the centre of the galaxy, summed over the 20\% brightest pixels of the galaxy \citep{lotz04}. 
\end{itemize}
Even though several of these parameters appear to be redundant, SVM were specially designed to be robust to redundancies in the feature space \citep[see e.g][]{huertas08}. The set of parameters used to obtain the morphological classification was tested in \cite{povic12}. Mentioned parameters were measured for both: local sample (redshifted and scaled in luminosity) and the ALH sample that we want to classify. Using vector machine, and comparison between the galSVM classification of local galaxies and their original, visual one, we can than classify ALH galaxies. The output of the classification step is a probability value to be in a given class for each object. The probability takes values from 0 to 1 (99.9 for unclassified objects). The robustness of the classification and the sensitivity to the training set is estimated by repeating the classification several times through Monte Carlo runs (hereafter MC) with slightly different training sets \citep[see][for more details]{huertas11}. We describe in the following the specific configuration used in this work.

\subsection{galSVM configuration and classification}

\subsubsection{Training sample of local galaxies}
\label{subsec_local_sample}

\begin{figure*}
\centering
\begin{minipage}[c]{.49\textwidth}
\includegraphics[width=8.35cm,angle=0]{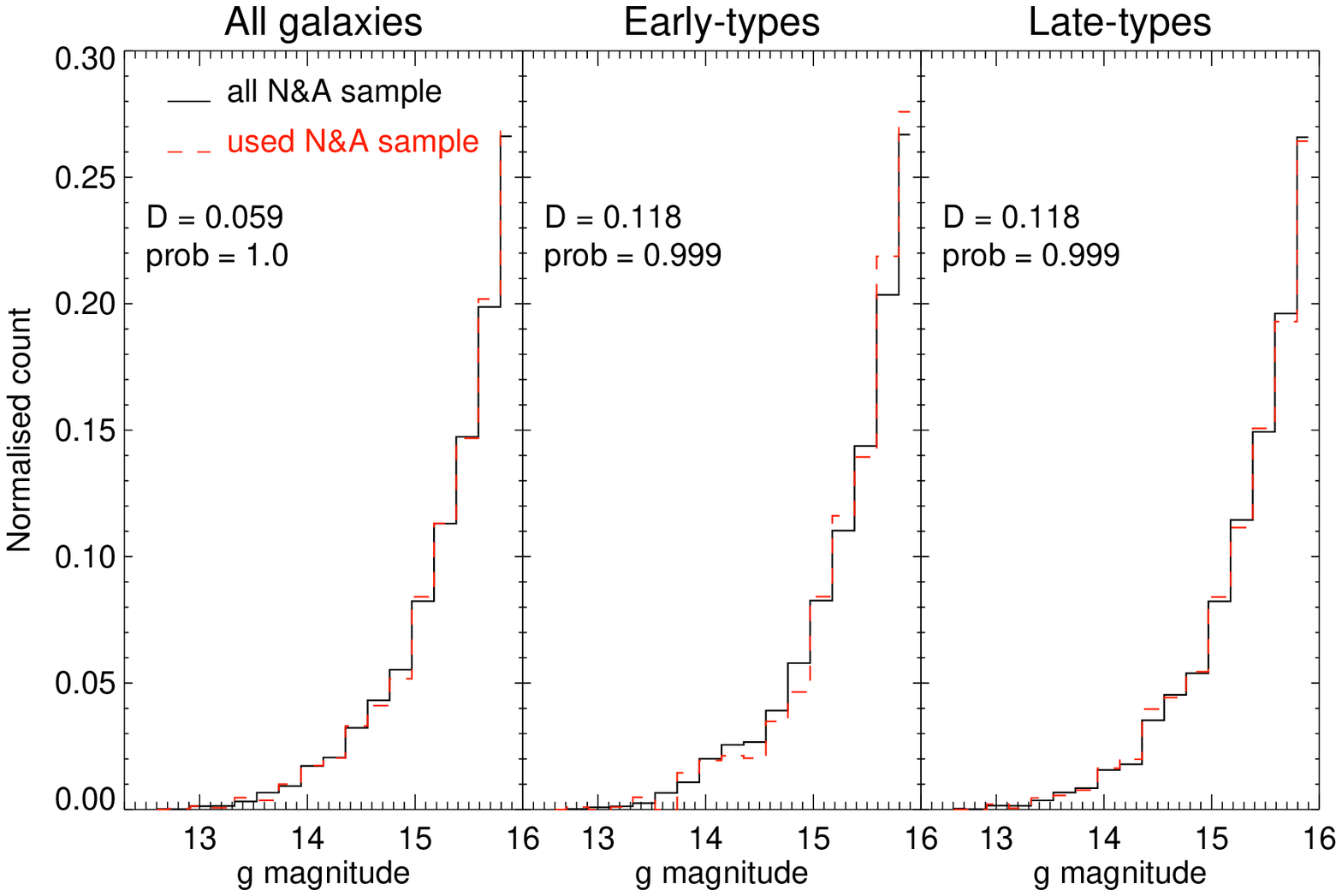}
\end{minipage}
\begin{minipage}[c]{.49\textwidth}
\includegraphics[width=8.35cm,angle=0]{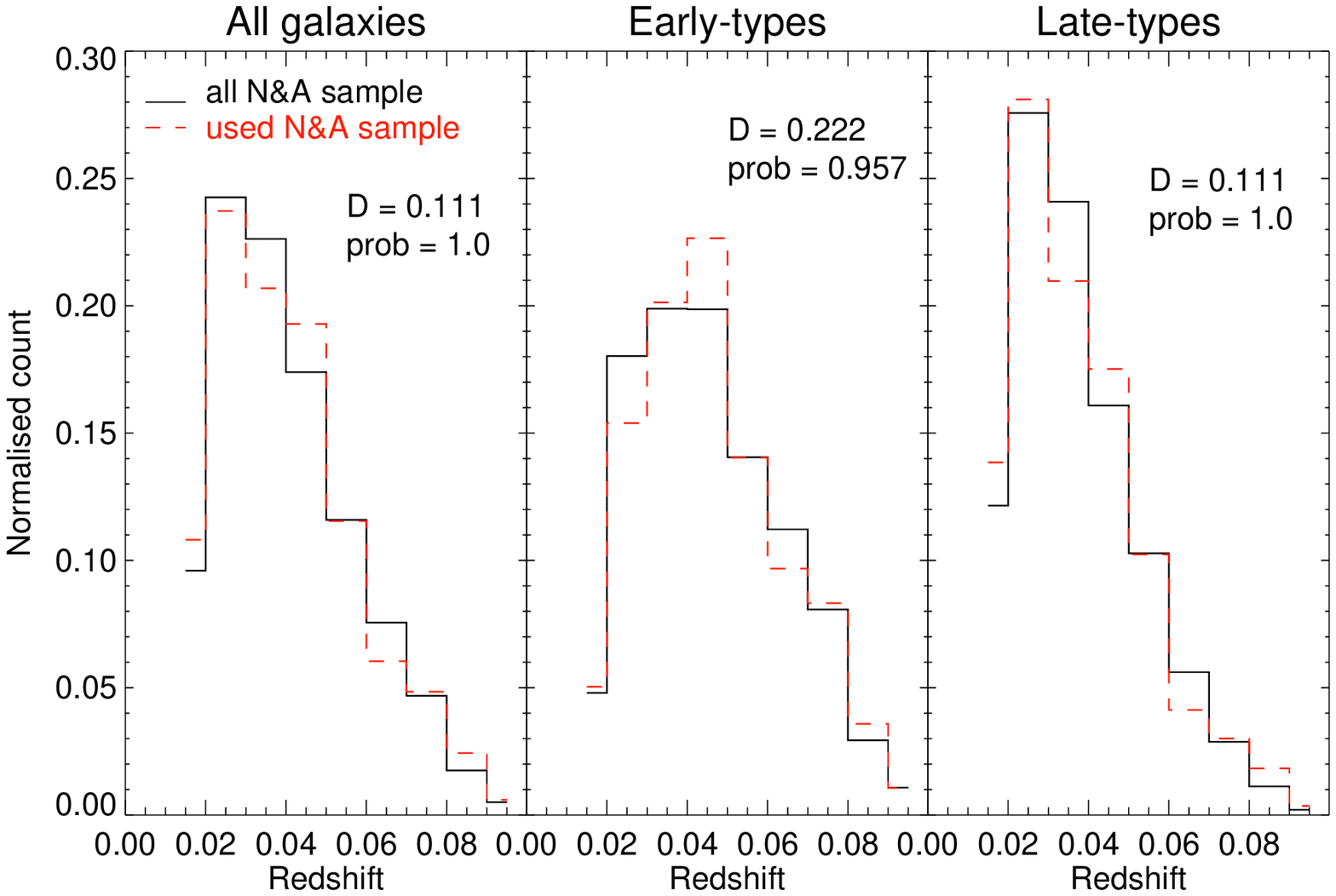}
\end{minipage}
\begin{minipage}[c]{.51\textwidth}
\includegraphics[width=8.35cm,angle=0]{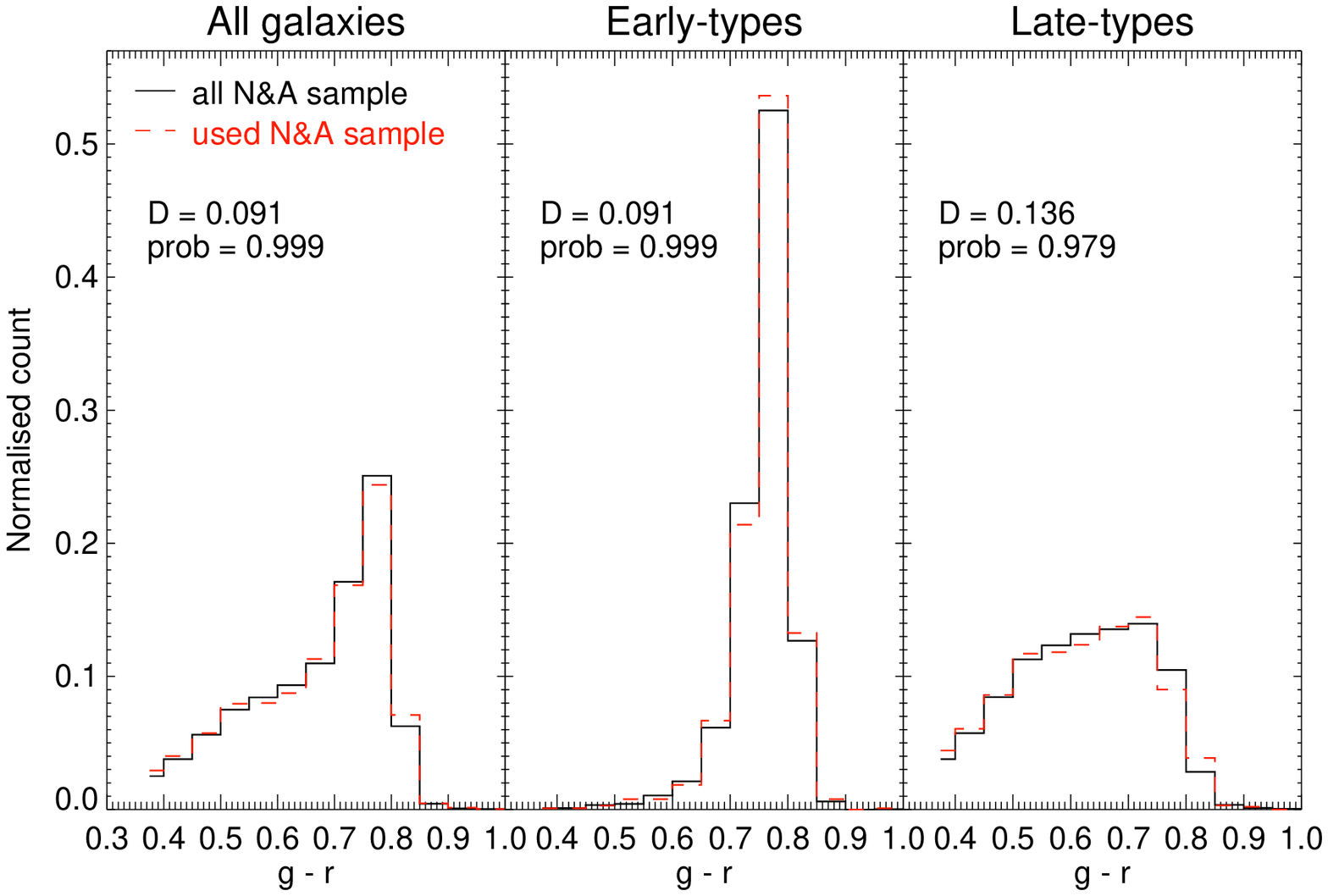}
\end{minipage}
\begin{minipage}[c]{.49\textwidth}
\includegraphics[width=8.35cm,angle=0]{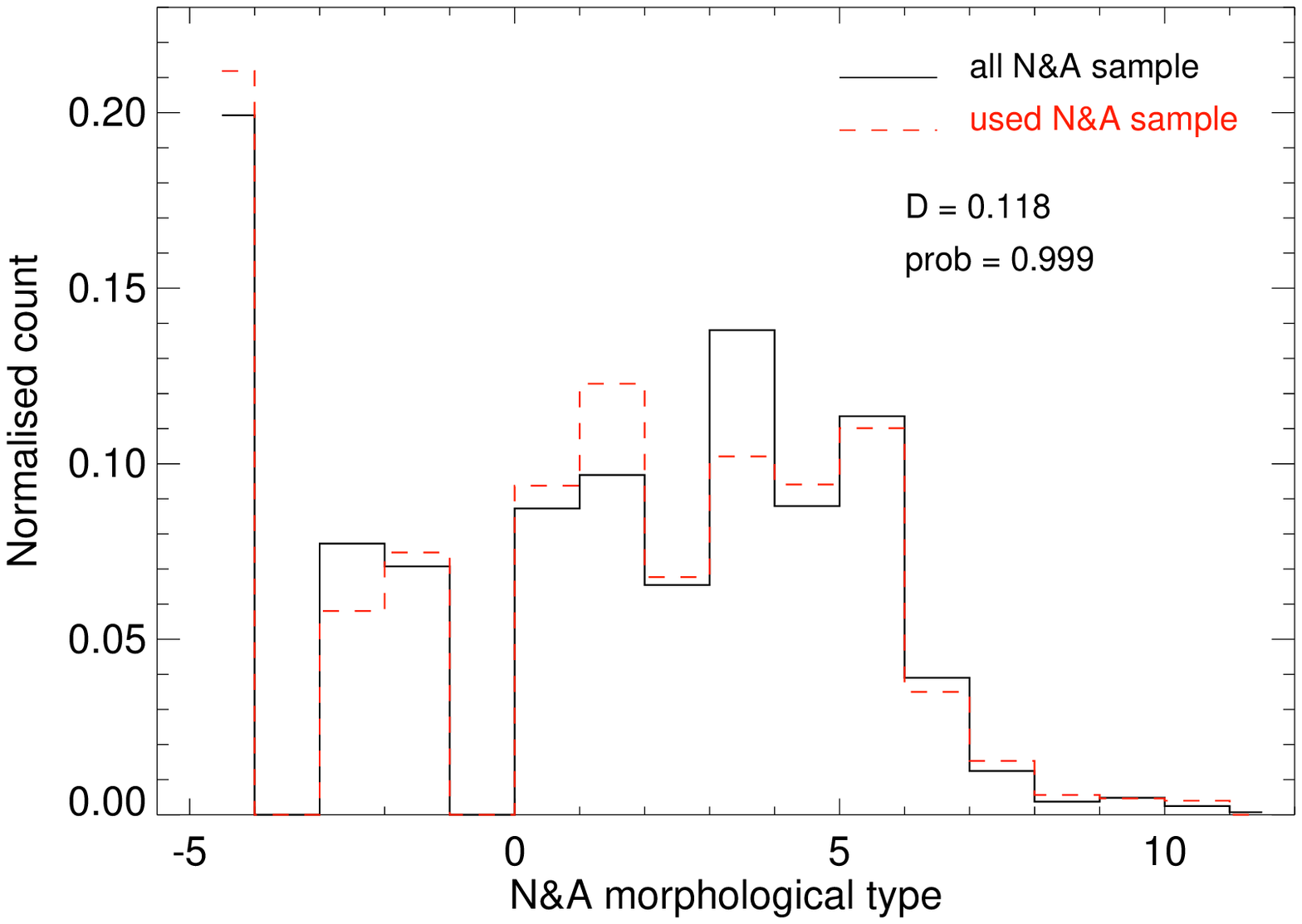}
\end{minipage}
\begin{minipage}[c]{.49\textwidth}
\includegraphics[width=8.35cm,angle=0]{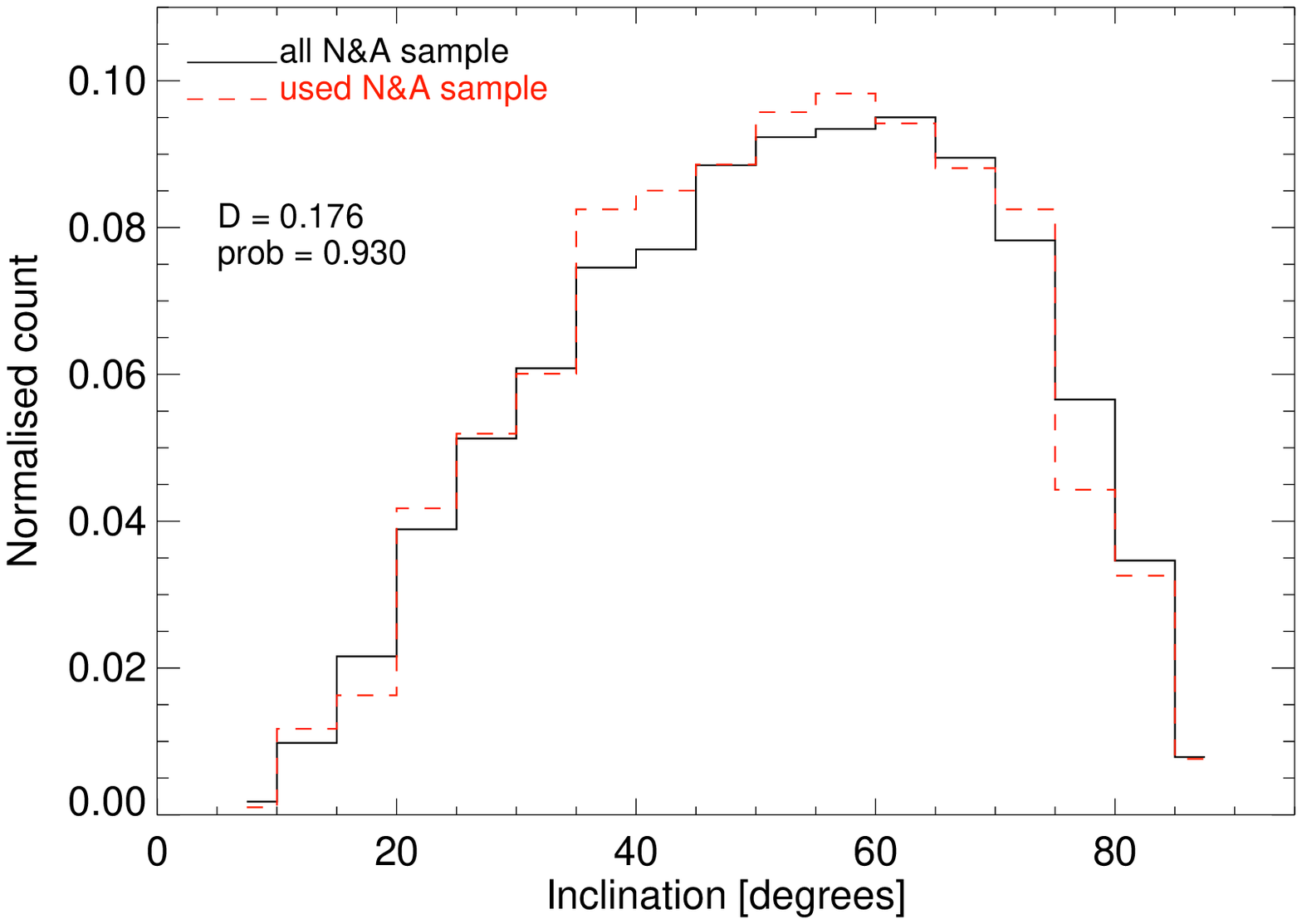}
\end{minipage}
\caption[ ]{Comparison between the properties of the full N\&A sample (black solid lines) and 3000 galaxies (red dashed lines) selected randomly from the full one to be used in our morphological classification. $g$ band magnitude \textit{(top, left)}, redshift \textit{(top, right)}, and $g$\,-\,$r$ colour \textit{(middle)} distributions are represented for all galaxies (left subdiagram), ETs (middle subdiagram), and LTs (right subdiagram). We also compared  morphological types (for all galaxies; \textit{bottom, left}) and inclinations (for late-type galaxies only; \textit{bottom, right}) distributions. Maximum deviation (D) and probability that two distributions are the same (prob; takes values between 0 and 1, where small values show that the cumulative distribution function of selected sample is significantly different from the full sample) are results of the KS statistics, showing in all plots that properties of our selected N\&A sample are completely consistent with the full one.}
\label{fig_na_all_3000_comparison}
\end{figure*}

\indent We use a local sample of 3,000 visually classified galaxies (0.01\,$\le$\,z\,$\le$\,0.1) taken from the Nair \& Abraham (2010; hereafter N\&A) catalogue which contains $\sim14.000$ galaxies from the SDSS. The number of galaxies used for training is selected as a trade-off between classification accuracy and computing time. The computing time to train the SVM with the current algorithm is indeed very sensitive to the size of the training data set. We therefore chose objects randomly and select the same number of early-type (elliptical and lenticular) and late-type (spiral and irregular) galaxies, to avoid problems related to unbalanced data sets, which is required for galSVM to work properly. We checked that the randomly selected subsample is representative in terms of general properties (colours, magnitudes, etc.) of the complete data set. Figure~\ref{fig_na_all_3000_comparison} shows these comparisons for the $g$ band magnitude, redshift, $g$\,-\,$r$ colour, N\&A morphological classification and inclination (for late-type galaxies only). Moreover, we also compared magnitudes, redshifts, and colours of ET and LT galaxies in both, selected and full samples. As can be seen from normalised distributions, the properties of randomly selected and full N\&A samples are consistent in all plots, as stated by Kolmogorov-Smirnov (hereafter KS) tests which proves that no selection biases are introduced. \\
\indent N\&A classification was obtained in the SDSS $g$ band images, while our classification was carried out in the F613W band (corresponds to SDSS $r$ band; see Sec.~\ref{subsec_galsvm_config_classifiaction}). Taking into account that we are performing broad classification, separating all sources into early- and late-type galaxies, the difference between the $g$ and $r$ bands inspected visually is insignificant for our work. For each morphological type from N\&A catalogue, we selected randomly 50 sources and we checked (between 4 people) their images in all 5 SDSS bands. In all cases we do not see any differences in galaxy structures between the $g$, $r$, and $i$ bands. On the other hand, we do see significant changes between any of these three bands and $u$ or $z$ filters.\\
\indent The selected local dataset is then redshifted, scaled in luminosity and dropped in the ALH fields. To that purpose, galSVM requires the apparent magnitude and photo-z distributions of all sources. To improve statistics, we used for each field the mag and photo-z distributions of the 4 CCDs after checking that these distributions are completely consistent with those of individual fields (individual CCDs). Before dropping the mock galaxies they are re-sampled with the same pixel scale than the ALH galaxies and convolved with a PSF to match the same spatial resolution. In order to cope with the k-correction we select the SDSS filter which is closer to the wavelength the ALH filter is probing given the redshift of the galaxy. Surface brightness dimming is taken into account when scaling the galaxy in flux since we empirically match the magnitude counts of the ALH survey (we do not introduce any size evolution though). Concerning the noise, we make the hypothesis that the noise from SDSS galaxies is negligible compared to the noise of galaxies at higher redshift (since we are using very bright galaxies). By dropping the galaxy in a real background we expect to reproduce at best all the different noises from the real images.\\
\indent \textbf{We assumed that there is no change in galaxy properties between the local and high-redshift samples (e.g. luminosity-morphology dependence). This assumption might be strong for our redshift range ($\sim$\,0\,-\,1), but can be justified since we are classifying all galaxies into two broad morphological types. Moreover, we only used morphological parameters in our classification, excluding luminosity. This minimises significantly the luminosity-morphology dependence, however it does not eliminate it completely (e.g., still higher redshift sample will contain more luminous sources, besides, for the redshifted training sample we are trying to reproduce the magnitude distribution of ALH sources assuming that luminosity/morphology relations are similar). To minimise even more the luminosity effects on our morphological classification, we forced the galSVM to select $\sim$\,50\% of early- and late-type local galaxies in each training MC run.}

\indent Although, we used 3,000 local galaxies in this work, a larger sample was constructed from the N\&A and EFIGI\footnote{http://www.astromatic.net/projects/efigis} \citep{baillard11} samples, with the total of 15,036 sources down to redshifts z\,$\le$\,0.1. The EFIGI visual classification was used at redshifts z\,$<$\,0.01, including all galaxies with good morphological classification, and excluding dwarf objects, while N\&A galaxies rang between 0.01\,$\le$\,z\,$\le$\,0.1. At the moment, this is the most complete local sample that can be used to test the morphology of high-redshift galaxies with the galSVM code, including for each source the list of astrometric, photometric, redshift, morphological type, size parameters, poststamps, masks, and PSF images in all five SDSS bands.

\subsubsection{galSVM classification}
\label{subsec_galsvm_config_classifiaction}

\begin{figure}
\centering
\includegraphics[width=0.49\textwidth,angle=0]{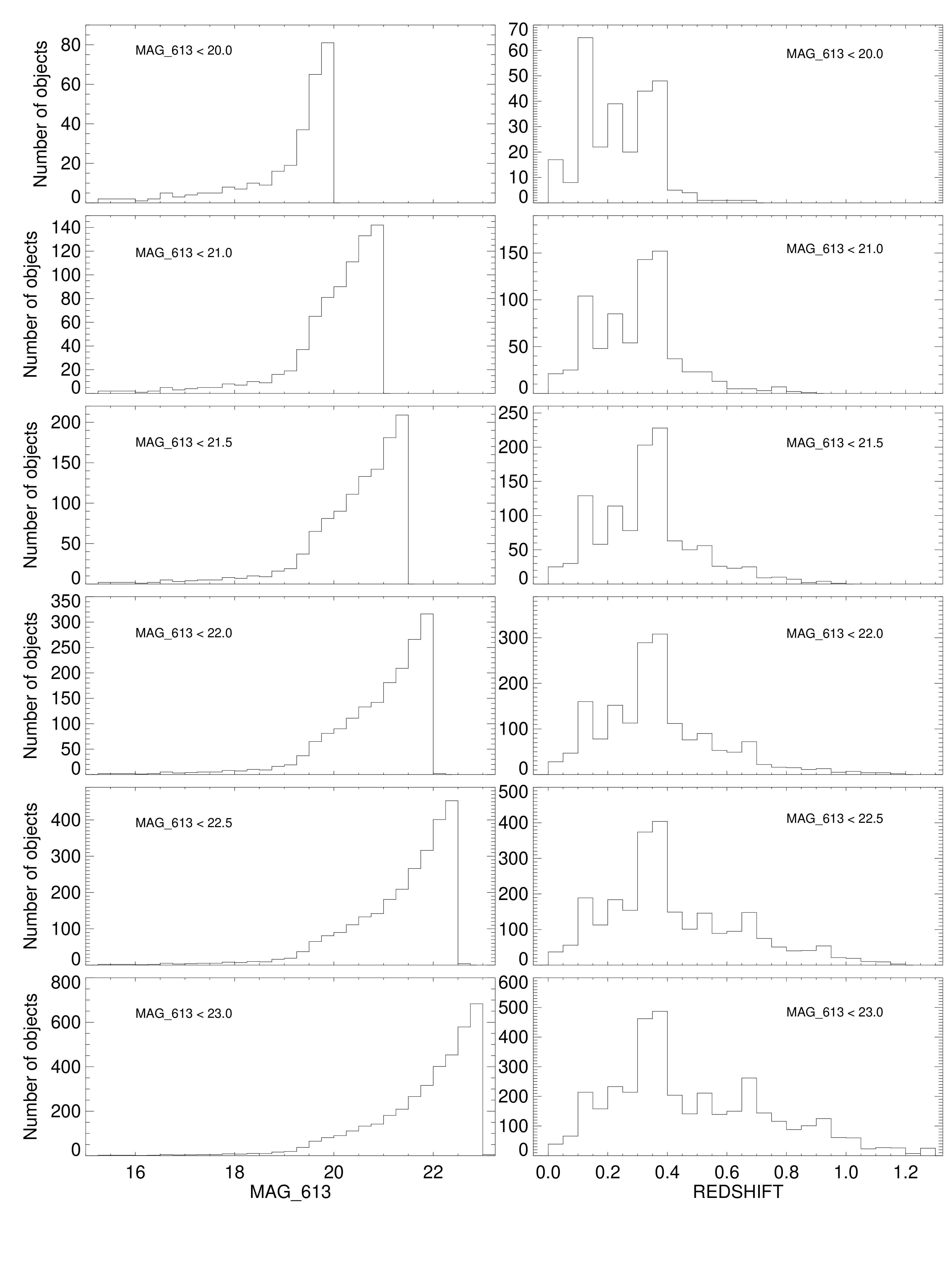}
\caption{\textit{(From top to bottom:)} Distributions of F613W magnitudes \textit{(left)} and corresponding redshifts \textit{(right)} in the ALH-4 field for six magnitude cuts used in our morphological classification: $\le$\,20.0, $\le$\,21.0, $\le$\,21.5, $\le$\,22.0, $\le$\,22.5, and $\le$\,23.0. 
\label{fig_mag_z_histos_alh4}}
\end{figure}

\indent The final classification is performed in the ALH F613W band, since the signal-to-noise ratio is higher in this filter, as showed in \cite{aparicio10}. For each galSVM run, we set the quality parameters that correspond to used ALH image: filter name, central wavelength, full width half maximum (FWHM), pixel scale, sigma, zero point, and saturation level. Given the typical resolution of the ALH survey (mainly around 1.0\,arcsec, see Table~\ref{tab_alhambra_obs}) and the depth, we have restricted in this work to two broad morphological classes: ET and LT. For each source in the final selected sample (see Sec.~\ref{subsec_real_sample}), we compute 7 morphological parameters described in Sec.~\ref{subsec_method}.\\
\indent Since the sample is fully dominated by faint objects given the shape of the magnitude counts (Fig.~\ref{fig_mag_z_histos_alh4}), when fainter objects are included in the classification, the algorithm will be optimized to classify these galaxies and the fraction of misclassified bright objects  might significantly increase. To avoid this effect, we performed the morphological classification with six increasing magnitude cuts: $\le$\,20.0, $\le$\,21.0, $\le$\,21.5, $\le$\,22.0, $\le$\,22.5, and $\le$\,23.0. That way, bright galaxies are classified using a training set made only of bright galaxies. These cuts also correspond roughly to increasing redshfits: $\le$\,0.7, $\le$\,0.9, $\le$\,1.0, $\le$\,1.2, $\le$\,1.2, and $\le$\,1.3. \\
\indent The final probability for each galaxy is computed as the average of the output probability of 15 MC independent runs. Again, this number is an empirical trade-off between computing time and accuracy, and is a result of previous tests carried out with 10, 15, and 20 runs. In each MC run, galSVM selects randomly a balanced set of 2,000 out of 3,000 input local galaxies. The result of each run is a probability for ALH galaxy to be ET, where for small probability values increases the possibility to be a LT. At the end, for each ALH source we compute the final probability to be ET (hereafter $p_{E}$) as average value of 15 probabilities, and the probability error as the scatter of the distribution. Obviously, since we deal with a 2-class problem only, the probability for the source to be LT (hereafter $p_{L}$) is simply $p_{L}=1-p_{E}$.  

\section{Calibration of the morphological classification using COSMOS/HST data}
\label{sec_class_reliability}

\indent The classification is calibrated in the ALH-4 field using the classification of the same objects observed with HST/ACS in the COSMOS survey. The morphological classification in  COSMOS used here was carried out with galSVM by \cite{huertas09}, and is publicly available through the \cite{tasca09} catalogue\footnote{http://irsa.ipac.caltech.edu/data/COSMOS/tables/morphology/\\cosmos\_morph\_tasca\_1.1.tbl}. The classification separates galaxies into three classes (early, late and irregulars) but does not have probabilistic information though since it was done with an older version of galSVM. \\
\indent By using COSMOS as the reference sample, we explicitly neglect the classification errors of the galaxies in COSMOS. This choice is justified since we focus here on bright galaxies with only two morphological classes. We also neglect the eventual morphological drift between the morphologies in COSMOS (F814W) and ALHAMBRA (F613W). Again,  since we are splitting our galaxies in two classes we do not expect a significant effect. This assumption is indeed confirmed through extensive tests on the SDSS dataset. We compared different morphological parameters of local N\&A galaxies between the $r$ and $i$ bands, without finding any significant differences in two bands, independently of the redshift and the Hubble type.\\

\begin{figure*}
\centering
\includegraphics[width=0.85\textwidth,angle=0]{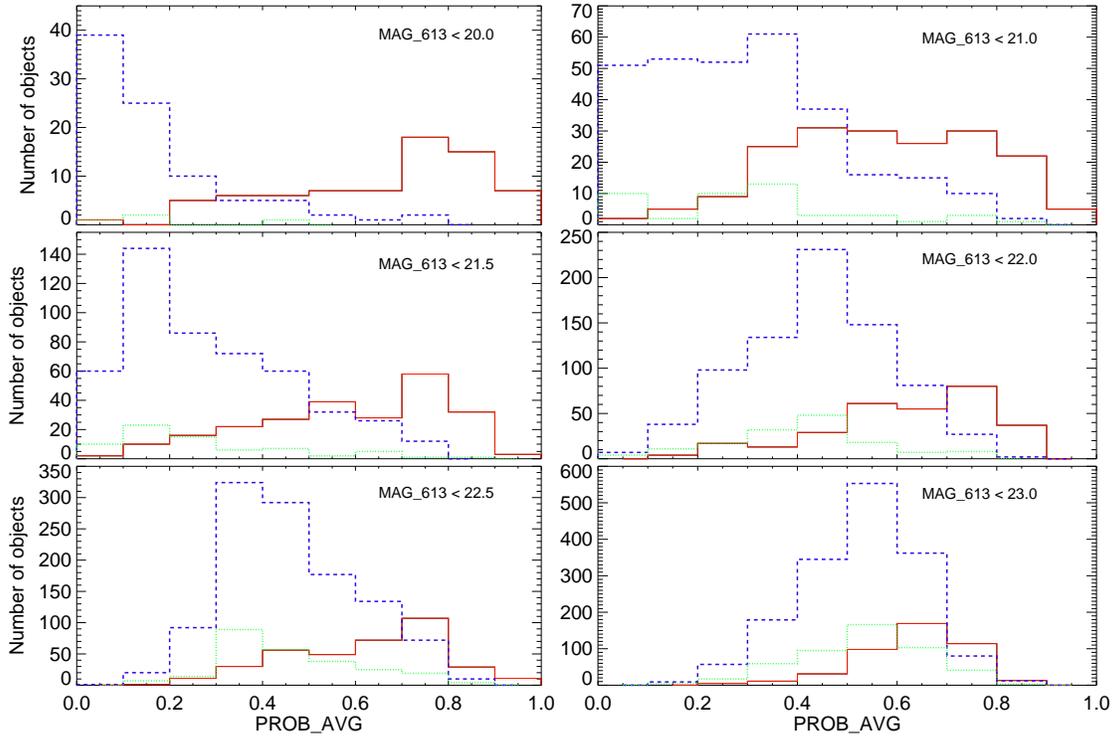}
\caption{Comparison between the galSVM classification in the ALH-4 and COSMOS surveys for different F613W magnitude cuts: $\le$\,20.0 \textit{(top, left)}, $\le$\,21.0 \textit{(top, right)}, $\le$\,21.5 \textit{(middle, left)}, $\le$\,22.0 \textit{(middle, right)}, $\le$\,22.5 \textit{(bottom, left)}, and $\le$\,23.0 \textit{(bottom, right)}. In each plot, we compare the distributions of averaged probabilities (PROB\_AVG) measured in the ALH-4 field for three morphological types classified in the COSMOS field using the HST/ACS imaging data \citep{huertas09,tasca09}: elliptical/S0 (red solid lines), spiral (blue dashed lines), and irregular (green dotted lines) galaxies.
\label{fig_cosmos_comparisons_flag0flag2}}
\end{figure*}

\indent Fig.~\ref{fig_cosmos_comparisons_flag0flag2} shows the distributions of $p_E$ derived in ALH-4 field for three morphological types classified in the COSMOS/HST survey with galSVM code: E/S0 (red solid lines), spirals (blue dashed lines), and irregulars (green dotted lines). Following these distributions, we look for the probability threshold ($p_{th}^E$,  $p_{th}^L$) to apply to the ALHAMBRA classification so that the resulting classification contains less than $10\% $ contaminations from the neighboring morphological type. To that purpose, for a given probability threshold $p_{th}^E$,$p_{th}^L$ we define the following parameters:
\begin{itemize}
\item {\it True positives (tp):} galaxies with $p_E>p_{th}^{E}$ ($p_L>p_{th}^{L}$) in ALH which are classified early-type (late-type) in COSMOS 
\item {\it True negatives (tn):} galaxies with $p_E<p_{th}^{E}$ ($p_L<p_{th}^{L}$) in ALH which are classified late-type (early-type) in COSMOS 
\item {\it False positives (fp):} galaxies with $p_E>p_{th}^{E}$ ($p_L>p_{th}^{L}$) in ALH which are classified late-type (early-type) in COSMOS 
\item {\it False negatives (fn):} galaxies with $p_E<p_{th}^{E}$ ($p_L<p_{th}^{L}$) in ALH which are classified early-type (late-type) in COSMOS 
\end{itemize}

The purity (P, fraction of well-classified objects among all objects classified in a given class) and the completeness (C, fraction of well-classified objects among all objects really belonging to a given class) are therefore defined as follows: 

\begin{equation}
\centering
P=1-\frac{fp}{fp+tp}
\end{equation}

\begin{equation}
C=\frac{tp}{fn+tp}
\end{equation}

$p_{10}^E$ and $p_{10}^L$ are therefore the thresholds to apply so that the contamination is lower than 10\% (P\,$>$\,0.9).  Table~\ref{tab_morph_reliability} shows the values obtained for the different magnitude cuts and for the 2 morphological classes. Notice that for ET galaxies fainter than 22, the contamination of LT galaxies is always above 10\%. \\
\indent Hereafter, a sample classified with $p_{10}$ thresholds we will call \textbf{p10 sample.} This is a sample that we release in paper and which we analyse in the following sections, as we explain in Sec.~\ref{sec_final_class}. However, in the same way we obtained the $p_{10}$ thresholds, we measured also the thresholds at higher contamination levels. We compare statistics in Sec.~\ref{sec_final_class}, while the full catalogue of whole $>$44,000 selected sources can be obtained through direct contact.\\
\indent Each probability threshold, has a corresponding completeness, which is reported later in Table~\ref{tab_final_class_number_cont}. For the p10 sample the completeness varies from $\sim70\%$ for the brightest objects to $\sim30\%$ at the faint end. These values become slightly worse if galaxies from CCD3 are included, since it is degraded compared to the others\footnote{http://www.caha.es/CAHA/Instruments/LAICA}. All statistics are carried out using only 'good detections', i.e. FLAG\,=\,0 (see Sec.~\ref{subsec_real_sample}). However, a similar accuracy is obtained when 'blended' (FLAG\,=\,2) sources are included. Therefore, in the following sections and in the published catalogue (Appendix~\ref{appendixA_cat}) we use both FLAG\,=\,0 and FLAG\,=\,2 sources. 

\noindent\begin{minipage}{\linewidth}
\centering \captionof{table}{$p_{10}$ threshold for selecting ET and LT galaxies (p10 sample) in different magnitude bins.
\label{tab_morph_reliability}}
\begin{tabular}{c | c c c c c c}
\hline
\textbf{Mag. limit}&\textbf{20.0}&\textbf{21.0}&\textbf{21.5}&\textbf{22.0}&\textbf{22.5}&\textbf{23.0}\\ 
\hline
\textbf{$p_{10}^{E}$}&$>$\,0.6&$>$\,0.7&$>$\,0.75&$>$\,0.8&--&--\\ 
\hline
\textbf{$p_{10}^{L}$}&$>$\,0.7&$>$\,0.7&$>$\,0.6&$>$\,0.5&$>$\,0.5&$>$\,0.5\\
\hline
\end{tabular}
\end{minipage}

\vspace{0.5cm}

\begin{figure*}
\centering
\includegraphics[width=0.55\textwidth,angle=0]{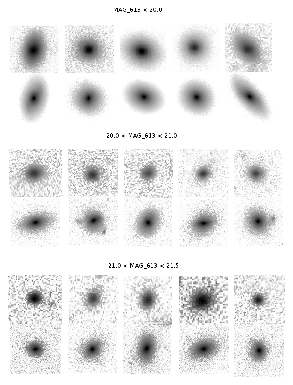}
\caption{Sample of ET galaxies in different magnitude bins, observed in the ALH-4 (top images, F613W band) and COSMOS/HST (bottom images, F814W band) surveys. 
\label{fig_images_et_alh_cosmos}}
\end{figure*}

\begin{figure*}
\centering
\includegraphics[width=0.99\textwidth,angle=0]{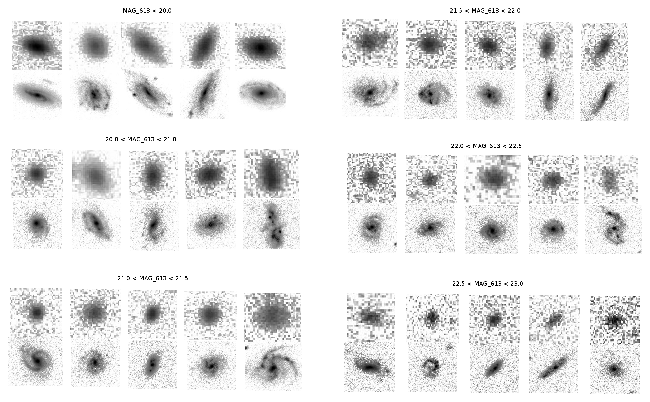}
\caption{Same as in Fig.~\ref{fig_images_et_alh_cosmos}, but for galaxies classified as LT.
\label{fig_images_lt_alh_cosmos}}
\end{figure*}

\indent As a sanity check, we performed a visual inspection of p10 classified sample in the ALH-4 field (overlaps with COSMOS). We used the HST/ACS images of all ETs (86 in total) and of 300 randomly selected LTs (20\% of the total p10 LT sample classified in the ALH-4 field). Visual classification was performed by five persons looking each object individually. By each classifier we measured the population of possible misclassified sources (within each type), and then we obtained the averaged one: 14.0\,$\pm$\,6.0\% and 7.8\,$\pm$\,4.8\% for ET and LT classified sources, respectively, which is in agreement with the statistics presented above. Figures~\ref{fig_images_et_alh_cosmos} and \ref{fig_images_lt_alh_cosmos} show a sample of selected ET and LT galaxy images, respectively, in different magnitude bins. For each ALH source (top images), we show also the corresponding HST/ACS image. 

\section{Morphological classification in ALHAMBRA}
\label{sec_final_class}

\indent After calibrating our morphology, we then run in the consistent way the galSVM code in whole ALH survey. In seven ALH fields, we set the galSVM configuration files, and for all detections we measured morphological parameters and averaged probabilities. Given the number of fields/observations and magnitude cuts (see Sec.~\ref{subsec_galsvm_config_classifiaction}), we obtained 288 catalogues. \\
\indent In total, with galSVM we obtained the classification for 85\% of selected sources (with different levels of contamination), while 15\% stayed unclassified. Sources for which we were able to measure morphological parameters have different levels of contamination by other morphological type. 
We analysed the unclassified sources in the ALH-4 field (overlaps with COSMOS) by using the HST/ACS images, detecting that $\sim$\,80\% of sample are interaction/merger candidates. In addition, $\sim$\,40\% of unclassified sources were detected with the CCD3, which in general has lower S/N in comparison with the other three CCDs (as already mentioned above). The population of ET/LT sources we might be missing for being unclassified with galSVM is about 3\%. \\
\indent For each magnitude bin we then applied the $p_{10}$ thresholds, obtained in the HST/ACS calibration phase to ensure a $10\%$ contamination level (p10 sample; see Table~\ref{tab_morph_reliability}). With this contamination, in all fields we keep classified 22,051 galaxies, 61\% of all classified sources, as showed in Table~\ref{tab_classified_contam} (table also provides the summary on other sources classified as ET/LT with higher contamination levels). Of those, 1,640 and 10,322 sources are classified as ET (photo-z\,$<$\,0.5) and LT (photo-z\,$<$\,1.0), respectively, down to magnitudes 22.0 in the F613W band. For this magnitude limit, the fraction of our ET galaxies respect to LTs is $\sim$\,16\% (consistent with the results obtained by \cite{huertas09} in the COSMOS field using the same method of galaxy classification). In addition, for magnitude range 22.0\,$<$\,F613W\,$\le$\,23.0 we classified other 10,089 LT galaxies with the same level of contamination of 10\% and down to photo-zs of 1.3. For the rest of the sample the contamination is higher (including also classified sources with very close companions (FLAG\,$\ge$\,3) which we excluded from the p10 sample), as showed in Table~\ref{tab_classified_contam}, being in general affected by the resolution of our data and by the worst quality of CCD3 data. Table~\ref{tab_final_class_number_cont} shows the number of objects and completeness of each morphological type in each magnitude bin. Beside the information about p10 sample, we also provide the statistics at higher contamination levels. As expected, the level of contamination is directly related with the magnitude, where fainter objects were classified with poorer probabilities. At higher contamination levels we have problems classifying ET sources, since many LTs start to mix with ETs giving worst probability distributions. For LT galaxies the contamination does not go above 20\%, so in general, if we are able to detect LT galaxy structures and classify the galaxy the contamination will always be low. When this is not the case, we have a contamination of ET sample with LT sources. Finally, for the p10 galaxies we expect to have a contamination by other type lower than 10\%, and contamination of stars below 1\% down to magnitudes 22.0 and 5\,-\,7\% between magnitudes 22.0 and 23.0. (see Sec.~\ref{subsec_real_sample}). This low-contamination catalogue is the one we are releasing with the present paper. Description of all columns and a sample of catalogue for the first five objects are available in Appendix~\ref{appendixA_cat}. The full p10 catalogue of 22,051 galaxies is available in the electronic version of this paper and through the ALHAMBRA website http://alhambrasurvey.com/.

\noindent\begin{minipage}{\linewidth}
\centering \captionof{table}{Population of sources classified as ET or LT with different levels of contamination by other type. \label{tab_classified_contam}}
\begin{tabular}{c | c}
\hline
\textbf{Contamination level}&\textbf{Population}\\ 
\hline
\textbf{$<$\,10\%}&\textbf{61\%}\\ 
$<$\,20\%&73\%\\ 
$<$\,30\%&74\%\\ 
$<$\,40\%&77\%\\ 
$<$\,50\%&82\%\\ 
$>$\,50\%&18\%\\ 
\hline
\end{tabular}
\end{minipage}

\begin{table*}
\begin{center}
\caption{Number and completeness of ET and LT galaxies, classified in each magnitude bin in relation with the contamination level. Beside each number we also represent the percentage of galaxies respect to the total number of sources in the observed contamination category (last column).
\label{tab_final_class_number_cont}}
\begin{tabular}{| c | c | c | c | c | c | c | c | c |}
\hline
&\textbf{F613W magnitude}&\textbf{$\le$\,20.0}&\textbf{20.0\,-\,21.0}&\textbf{21.0\,-\,21.5}&\textbf{21.5\,-\,22.0}&\textbf{22.0\,-\,22.5}&\textbf{22.5\,-\,23.0}&\\ 
\hline
&\textbf{Contamination\,$\le$\,10\%}&\textbf{715 (44\%)}&\textbf{549 (33\%)}&\textbf{219 (13\%)}&\textbf{157 (9\%)}&\textbf{--}&\textbf{--}&\textbf{1640}\\
&\textbf{C$_{<10}$}&\textbf{0.7}&\textbf{0.4}&\textbf{0.4}&\textbf{0.3}&\textbf{--}&\textbf{--}&\\
\cline{2-9}
&\textbf{Contamination\,$\le$\,20\%}&1103 (28\%)&1841 (47\%)&622 (16\%)&252 (6\%)&112 (3\%)&--&3930\\
\textbf{ET}&\textbf{C$_{<20}$}&0.7&0.4&0.4&0.3&0.1&--&\\
\cline{2-9}
&\textbf{20\,$<$\,Contamination\,$\le$\,50\%}&--&--&706 (21\%)&1609 (47\%)&1081 (32\%)&--&3396\\
&\textbf{C$_{20-50}$}&--&--&0.6&0.5&0.4-0.5&--&\\
\cline{2-9}
&\textbf{Contamination\,$>$\,50\%}&--&--&--&492 (7\%)&1877 (28\%)&4297 (65\%)&6666\\
&\textbf{C$_{>50}$}&--&--&--&0.7&0.6-0.7&0.8&\\
\hline
\hline
&\textbf{Contamination\,$\le$\,10\%}&\textbf{1082 (5\%)}&\textbf{2818 (14\%)}&\textbf{2689 (13\%)}&\textbf{3733 (18\%)}&\textbf{4828 (24\%)}&\textbf{5261 (26\%)}&\textbf{20411}\\
&\textbf{C$_{<10}$}&\textbf{0.7}&\textbf{0.7}&\textbf{0.7}&\textbf{0.7}&\textbf{0.6}&\textbf{0.3}&\\
\cline{2-9}
&\textbf{Contamination\,$\le$\,20\%}&1720 (7\%)&3432 (15\%)&3032 (13\%)&3659 (16\%)&4966 (22\%)&6177 (27\%)&22986\\
\textbf{LT}&\textbf{C$_{<20}$}&0.7&0.7&0.7&0.7&0.6&0.3&\\
\cline{2-9}
&\textbf{20\,$<$\,Contamination\,$\le$\,50\%}&--&--&--&--&--&--&\\
&\textbf{C$_{20-50}$}&--&--&--&--&--&--&\\
\cline{2-9}
&\textbf{Contamination\,$>$\,50\%}&--&--&--&--&--&--&\\
&\textbf{C$_{>50}$}&--&--&--&--&--&--&\\
\hline
\end{tabular}
\end{center}
\end{table*}

\vspace{0.5cm}

\indent We do not treat mergers specifically in our classification. However, we do minimise their population by excluding from our p10 classified sample all sources with SExtractor FLAG parameter $>$\,2 (include objects with close neighbours). They do enter in the total selected sample, but we excluded them from all analysis presented in this paper (see Sec.~\ref{sec_class_reliability}). Moreover, we saw that many interacting systems stayed unclassified with galSVM. As mentioned above, $\sim$\,80\% of all unclassified sources show clear signs of distortions, close companions, are edge-on systems, or a mixture. Finally, we checked the HST/ACS images of all p10 galaxies that overlap with the COSMOS field, in order to quantify the population of possible interactions. From independent inspections carried out by five persons, we obtained the averaged population of 15.0\,$\pm$\,2.5\% merger/interaction candidates between the galaxies classified as LT.   

\subsection{Selection effects}
\label{subsec_sel_effects}
Since the final catalogue is obtained after applying arbitrary probability cuts, it might be affected by non-trivial selection effects which we investigate in the following from the point of view of redshift and size distributions. 

\begin{figure*}
\centering
\includegraphics[width=0.75\textwidth,angle=0]{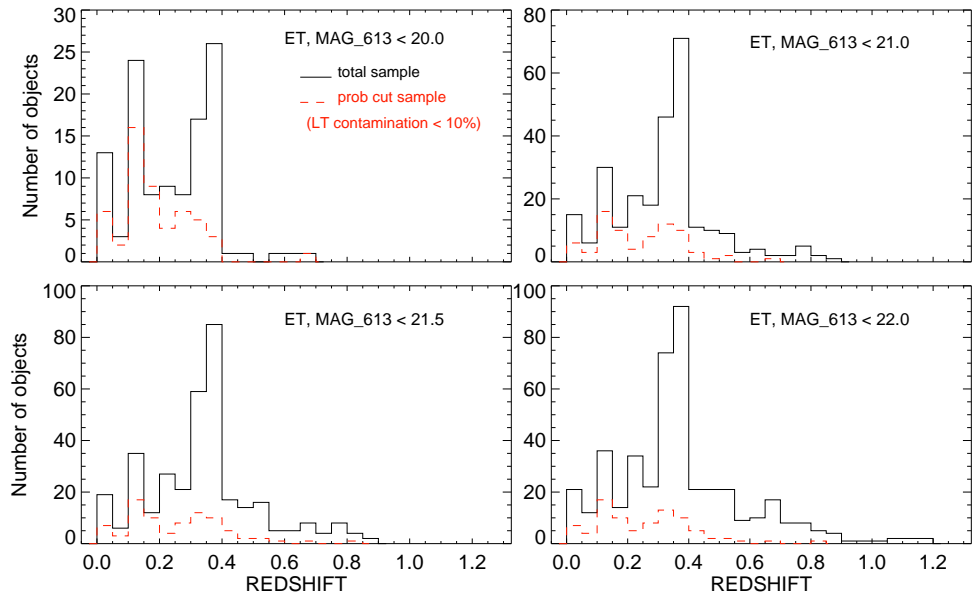}
\caption{Redshift distributions of total ET samples (black solid lines), selected through the magnitude cuts defined in Sec.~\ref{subsec_galsvm_config_classifiaction}, and of $p_{10}$ morphologically classified ET sample (red dashed lines), selected after applying the probability cuts obtained in the COSMOS/HST comparisons. Distributions are compared for four magnitude cuts, since above 21.5\,-\,22.0 the contamination of LT objects is too significant. On the other side, for LT galaxies (Fig.~\ref{fig_zdist_totsample_probcutsample_lt}) we compare the redshift distributions for all six magnitude cuts. 
\label{fig_zdist_totsample_probcutsample_et}}
\end{figure*}

\begin{figure*}
\centering
\includegraphics[width=0.85\textwidth,angle=0]{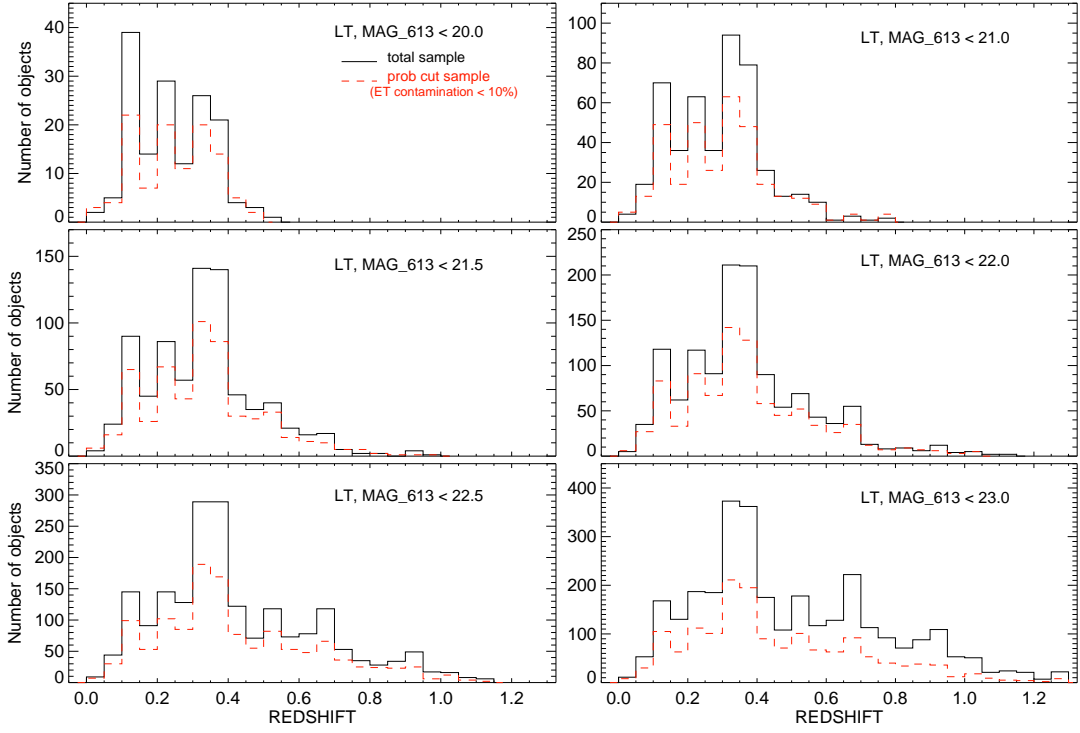}
\caption{Same as Fig.~\ref{fig_zdist_totsample_probcutsample_et}, but for LT galaxies observed in 6 magnitude bins.
\label{fig_zdist_totsample_probcutsample_lt}}
\end{figure*}

\begin{figure*}
\centering
\begin{minipage}[c]{.49\textwidth}
\includegraphics[width=0.99\textwidth,angle=0]{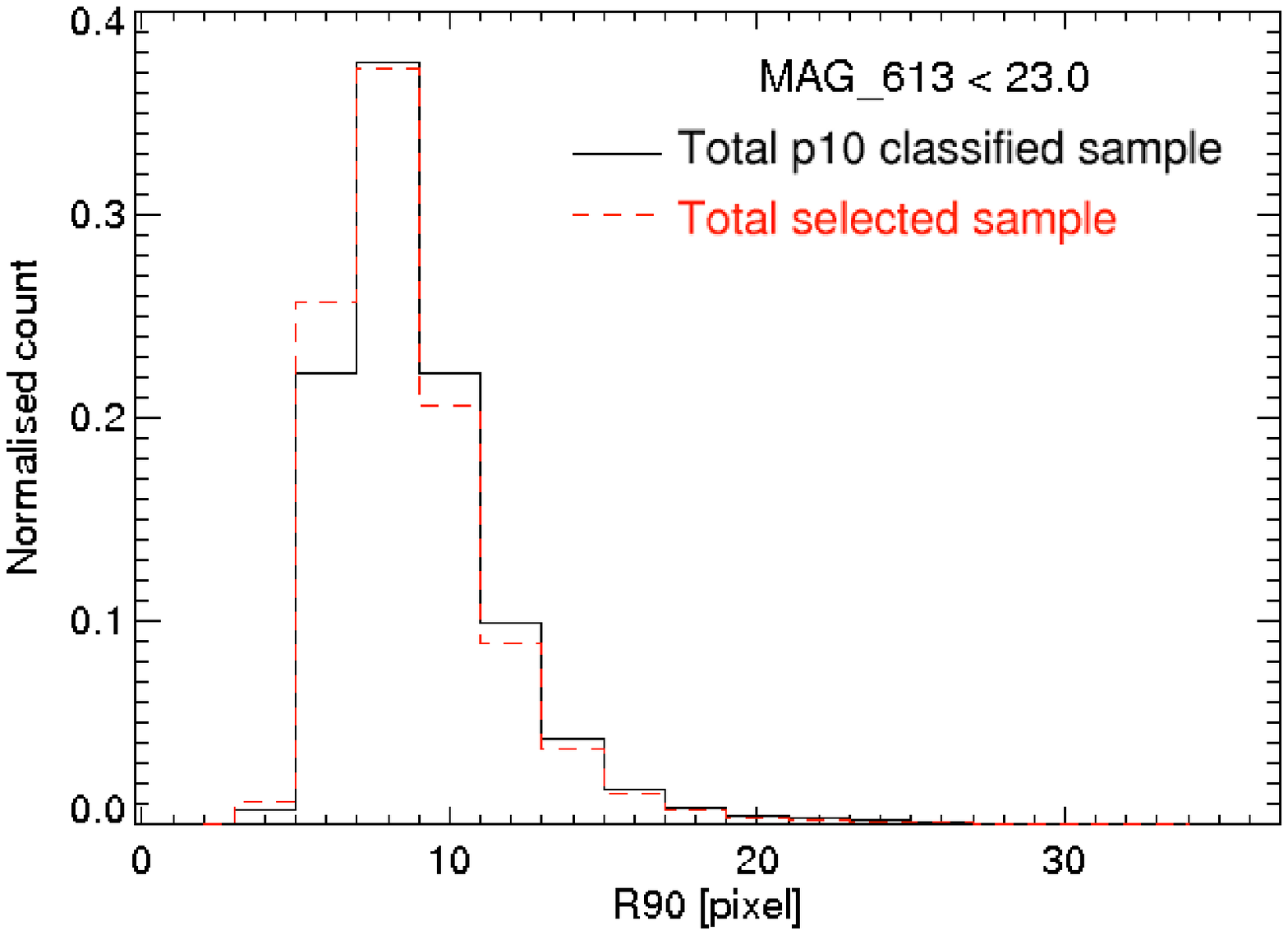}
\end{minipage}
\begin{minipage}[c]{.49\textwidth}
\includegraphics[width=0.99\textwidth,angle=0]{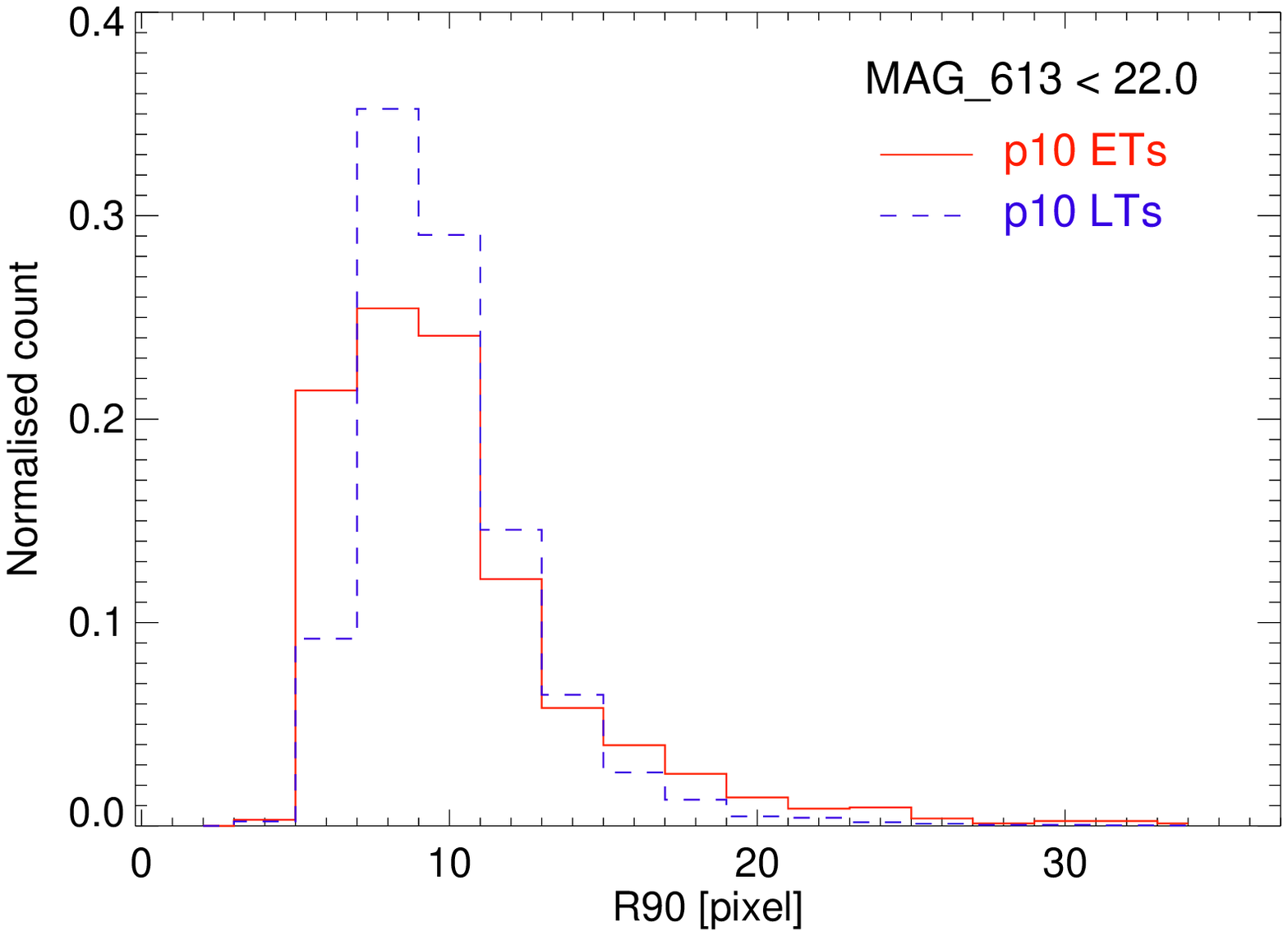}
\end{minipage}
\caption[ ]{Normalised distributions of radius at 90\% of flux for: total selected (red dashed lines) and $p_{10}$ classified (black solid lines) samples (left panel), and ET (red solid lines) and LT (blue dshed lines) galaxies (right panel).     
\label{fig_size_analysis}}
\end{figure*}

\begin{itemize}
\item \textit{\textbf{Redshift.}} Figures~\ref{fig_zdist_totsample_probcutsample_et} and ~\ref{fig_zdist_totsample_probcutsample_lt} show for each analyzed magnitude bin, the redshift distributions of ET and LT galaxies, respectively, of the full sample as compared to the final $p_{10}$ sample defined in the previous section. For a total sample (black solid lines), the classification is directly obtained from the \cite{tasca09} catalogue \citep[classification described in][]{huertas09}, putting all their 'ellipticals' as ET, and spirals and irregulars as LT. Sample marked with red dashed lines represent ALH $p_{10}$ sample, obtained through COSMOS/HST comparisons (criteria summarised in Table~\ref{tab_morph_reliability}). The redshift distributions of the late-type galaxies are consistent in both samples which indicates that there is no redshift selection bias. However, for ET galaxies, the redshift distribution of the $p_{10}$ sample peaks at lower values (0.2\,-\,0.4) than the full sample. This reflects the fact that early-type galaxies at high redshift cannot be identified properly because we start being seriously affected by the spatial resolution. The reduction of contamination therefore results in a sample of early-type galaxies with $z<0.5$. We will discuss this selection bias in terms of stellar mass and absolute magnitudes in Sec.~\ref{subsec_stm_mag_z}. \\
\item \textit{\textbf{Size.}} Figure~\ref{fig_size_analysis} (left panel) shows the comparison between the normalised distributions of galaxy size at 90\% of flux for the total selected sample of $>$\,40,000 sources (see Sec.~\ref{subsec_real_sample}) and total $p_{10}$ classified sample of 22,051 galaxies (see above), down to magnitudes 23.0 in the F613W band. Size distributions of ET and LT galaxies classified with $<$\,10\% contamination are showed in the right panel of the same figure, down to magnitudes 22.0 in the F613W band (where we still have classification into both types). We can confirm that down to the magnitude limit of 23.0 studied in this paper, we do not find any significant influence of galaxy size onto the morphological classification. KS statistic shows that the size distributions of total selected and $p_{10}$ ET and LT classified samples are consistent, having the KS probability parameter of 0.9994 (out of 1.0).  
\end{itemize}

\subsection{Properties of ET sample}
\label{subsec_et_sample}

\begin{figure*}
\centering
\begin{minipage}[c]{.49\textwidth}
\includegraphics[width=0.99\textwidth,angle=0]{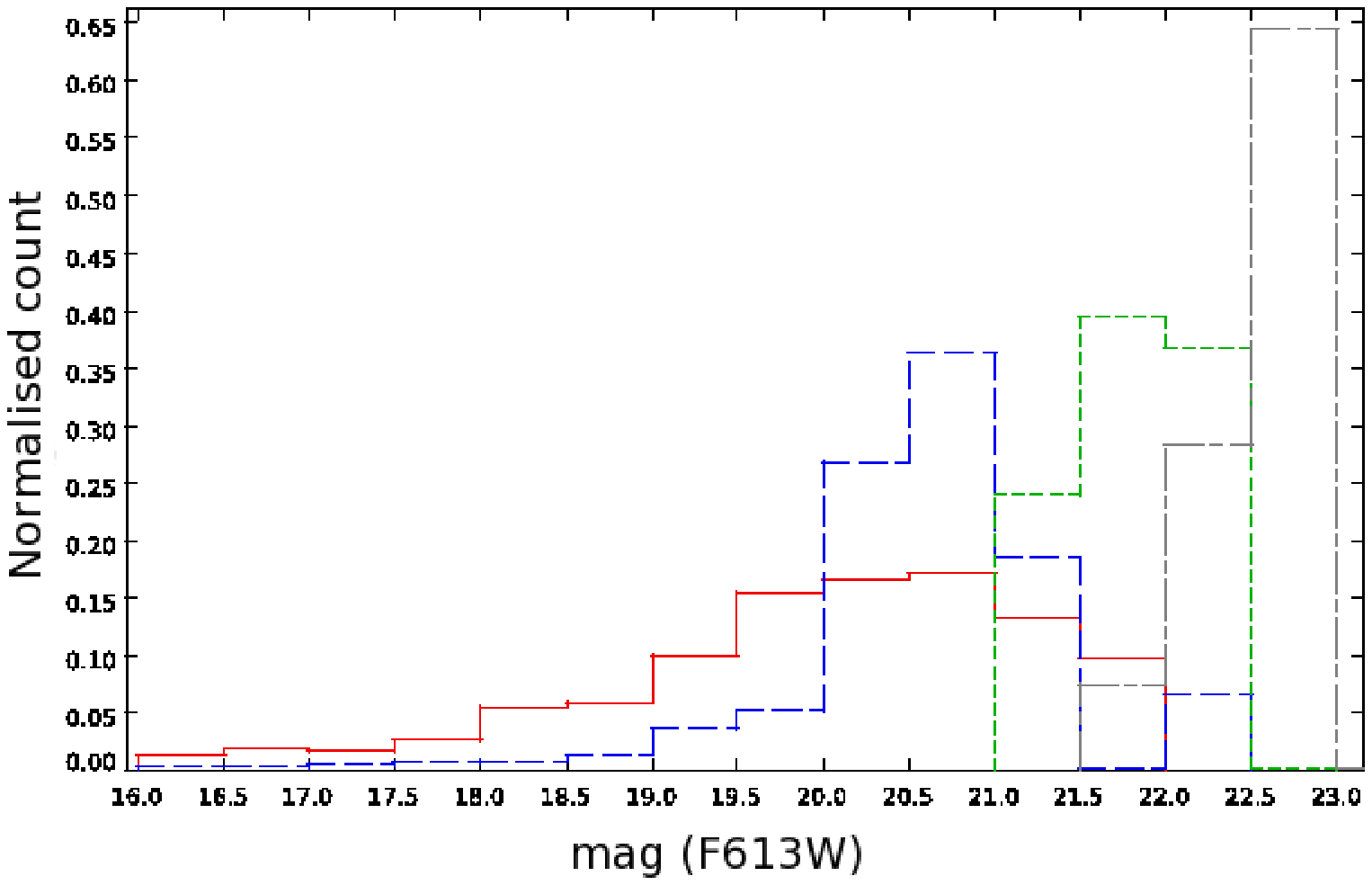}
\end{minipage}
\begin{minipage}[c]{.49\textwidth}
\includegraphics[width=0.99\textwidth,angle=0]{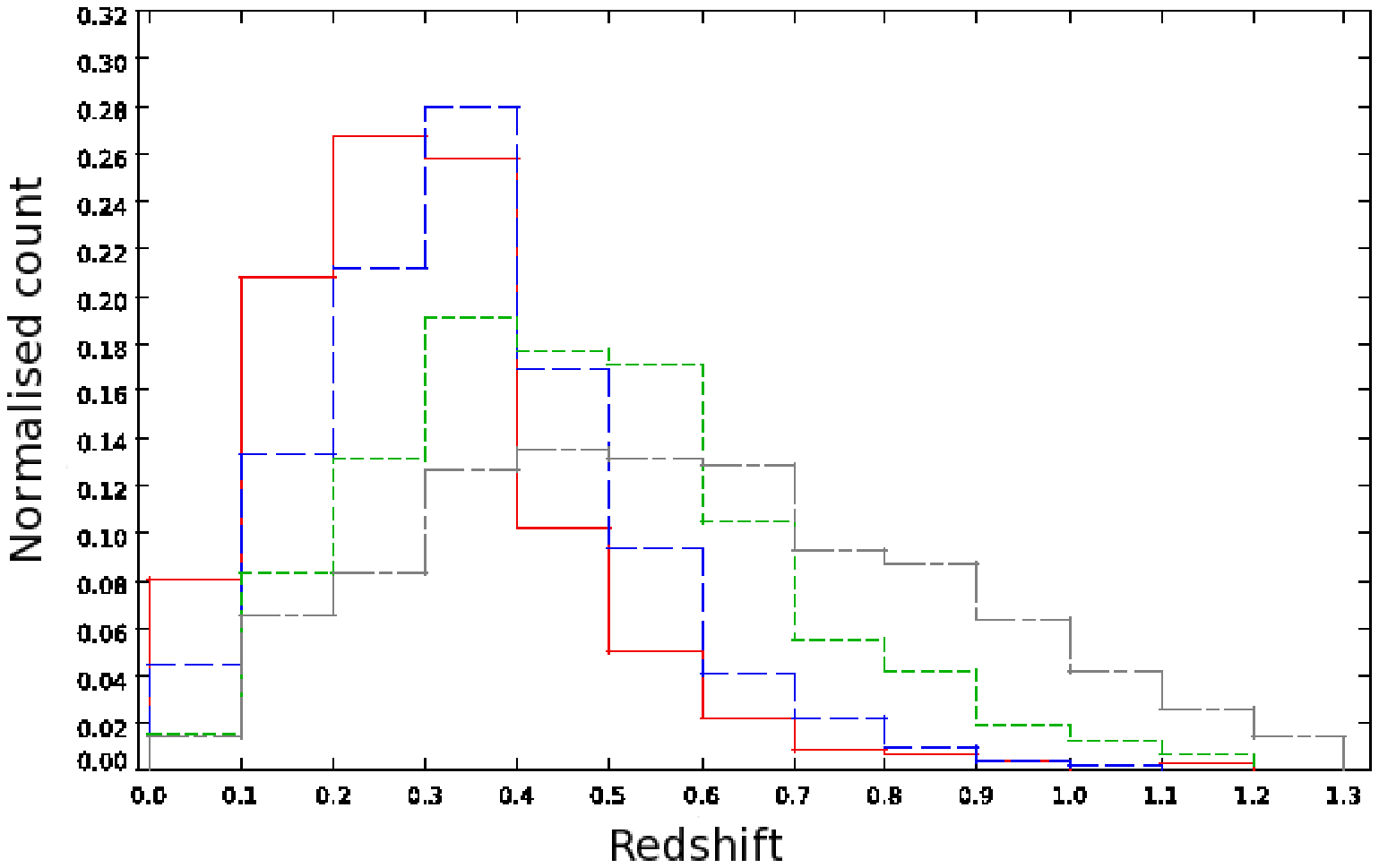}
\end{minipage}
\begin{minipage}[c]{.49\textwidth}
\includegraphics[width=0.99\textwidth,angle=0]{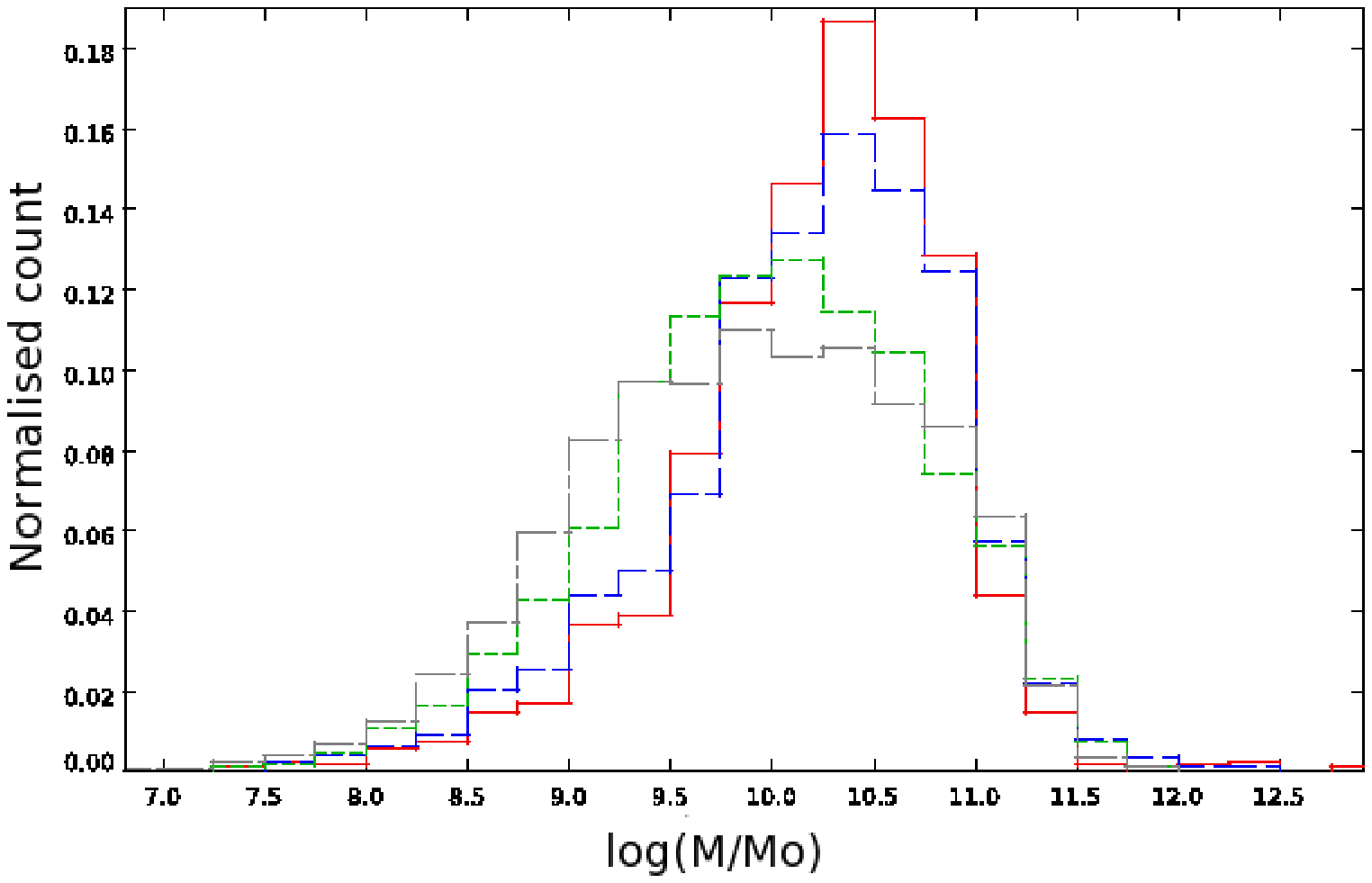}
\end{minipage}
\begin{minipage}[c]{.49\textwidth}
\includegraphics[width=0.99\textwidth,angle=0]{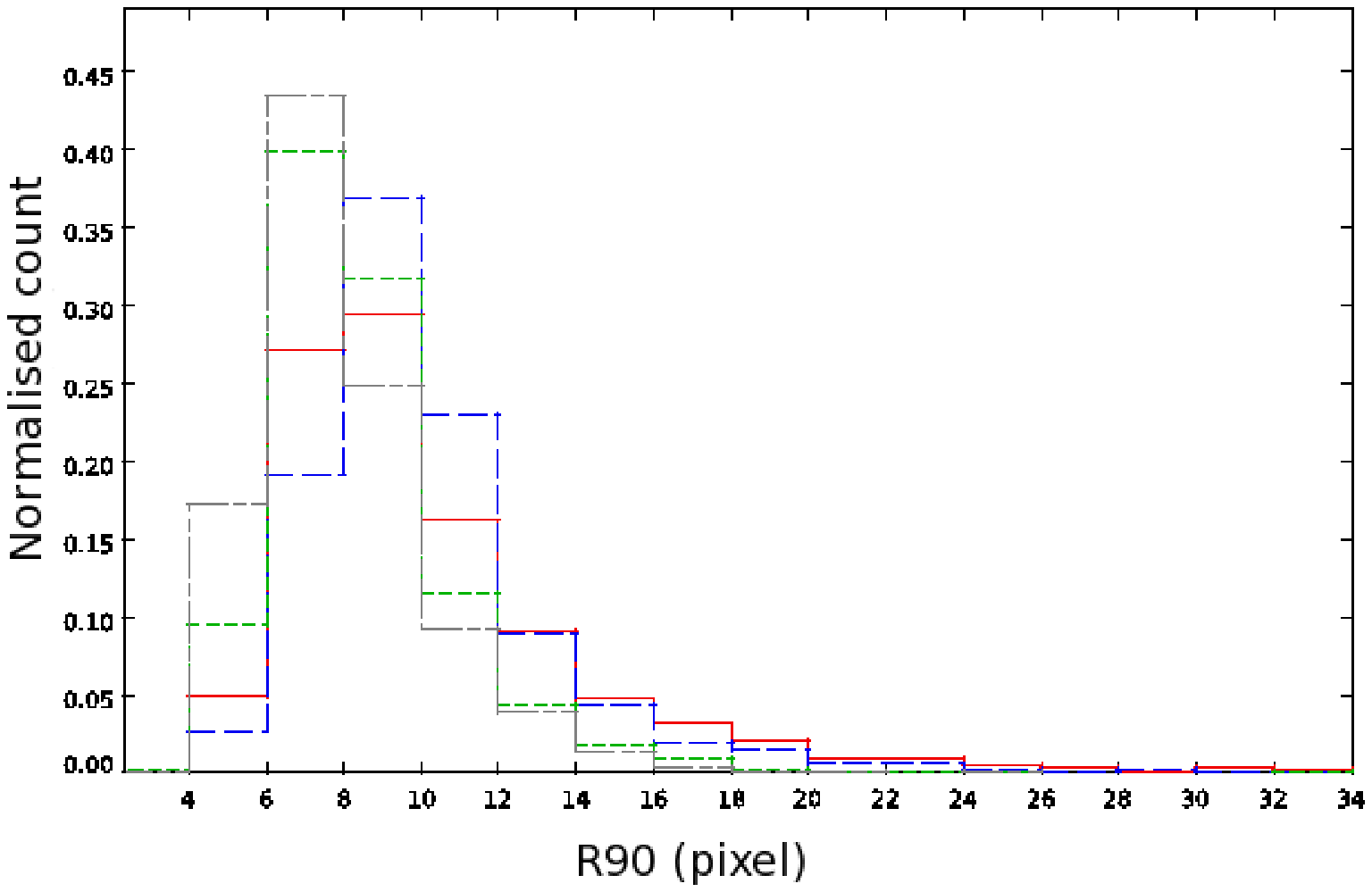}
\end{minipage}
\caption[ ]{Normalised distributions of F613W apparent magnitude \textit{(top, left)}, redshift \textit{(top, right)}, stellar mass \textit{(bottom, left)}, and radius at 90\% of flux \textit{(bottom, right)} of ET galaxies classified with different levels of contamination: $\le$\,10\% (red solid line), 10\,-\,20\% (blue dashed line), 20\,-\,50\% (green dotted line), and $>$\,50\% (grey dash-dot-dash line).    
\label{fig_et_samlpe_diffcontamlevels}}
\end{figure*}

\indent We compared different properties between the p10 ET sample and galaxies classified as ET with higher levels of contamination by LT type. Figure~\ref{fig_et_samlpe_diffcontamlevels} represents these comparisons, and shows the normalised distributions of apparent F613W magnitude, redshift, stellar mass, and radius at 90\% of flux. As can be seen, galaxies classified as ET with higher contamination levels ($>$\,20\%) are in general fainter and more distant, and have higher population of sources with lower stellar masses and sizes in comparison with the p10 ET sample. We performed the KS statistical test comparing (for all properties) p10 with other ET classified samples, obtaining in all cases that distributions are significantly different. Using t-means statistical test, for all sources, except when comparing stellar mass and size between p10 and 10\%\,$<$\,Contamination\,$\le$\,20\% samples, we found that p10 and other samples have significantly different means.

\section{Main properties of the morphological catalogue}
\label{sec_prop}
\indent In this Section we test the general properties of p10 galaxies classified as ETs or LTs and check that they populate the expected regions in classical relations (e.g., morphological diagrams, colour-magnitude and colour-stellar mass relations).

\subsection{Morphological parameters}
\label{subsec_morphparam}

\begin{figure*}
\centering
\begin{minipage}[c]{.49\textwidth}
\includegraphics[width=0.85\textwidth,angle=0]{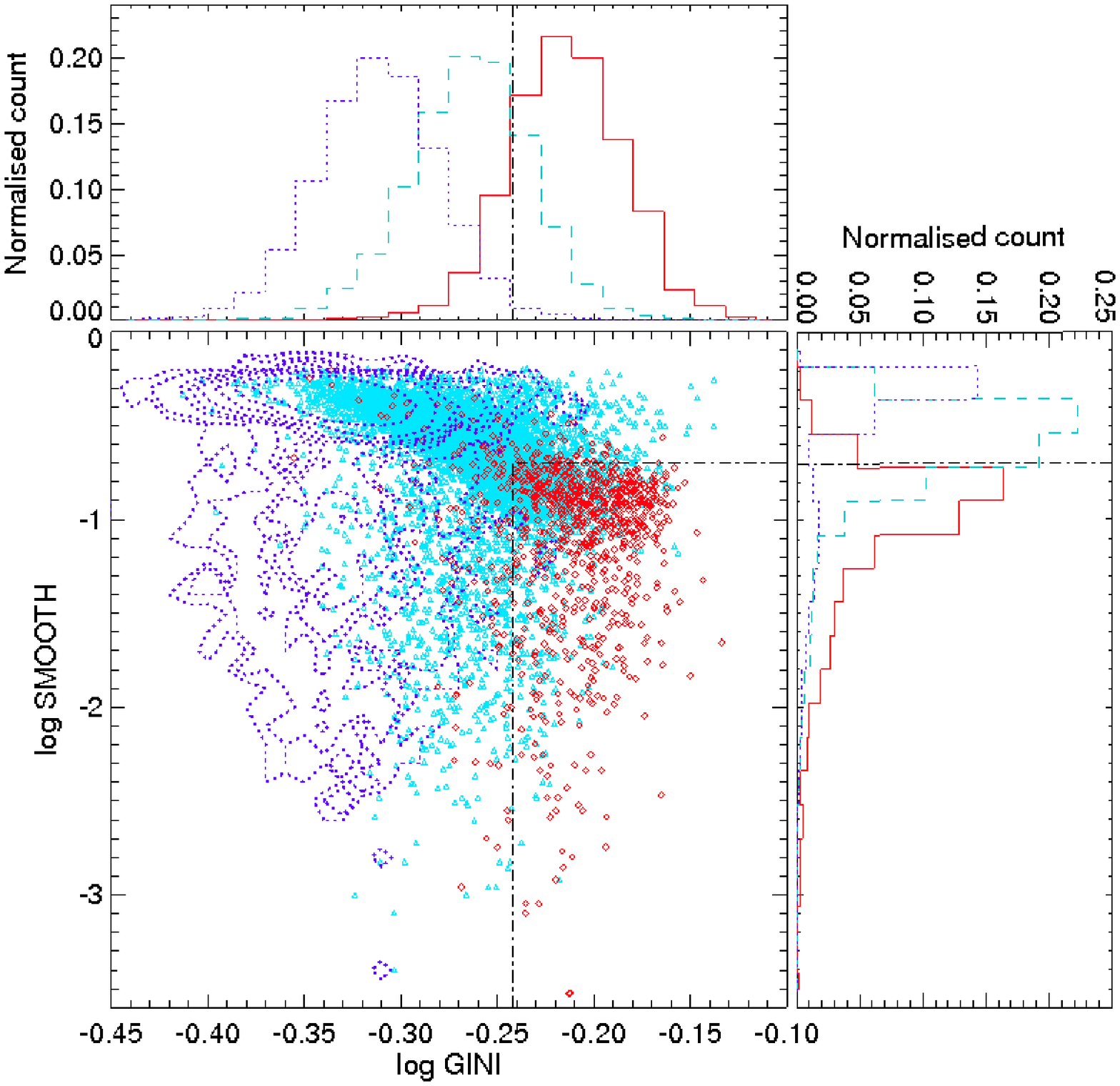}
\end{minipage}
\begin{minipage}[c]{.49\textwidth}
\includegraphics[width=0.85\textwidth,angle=0]{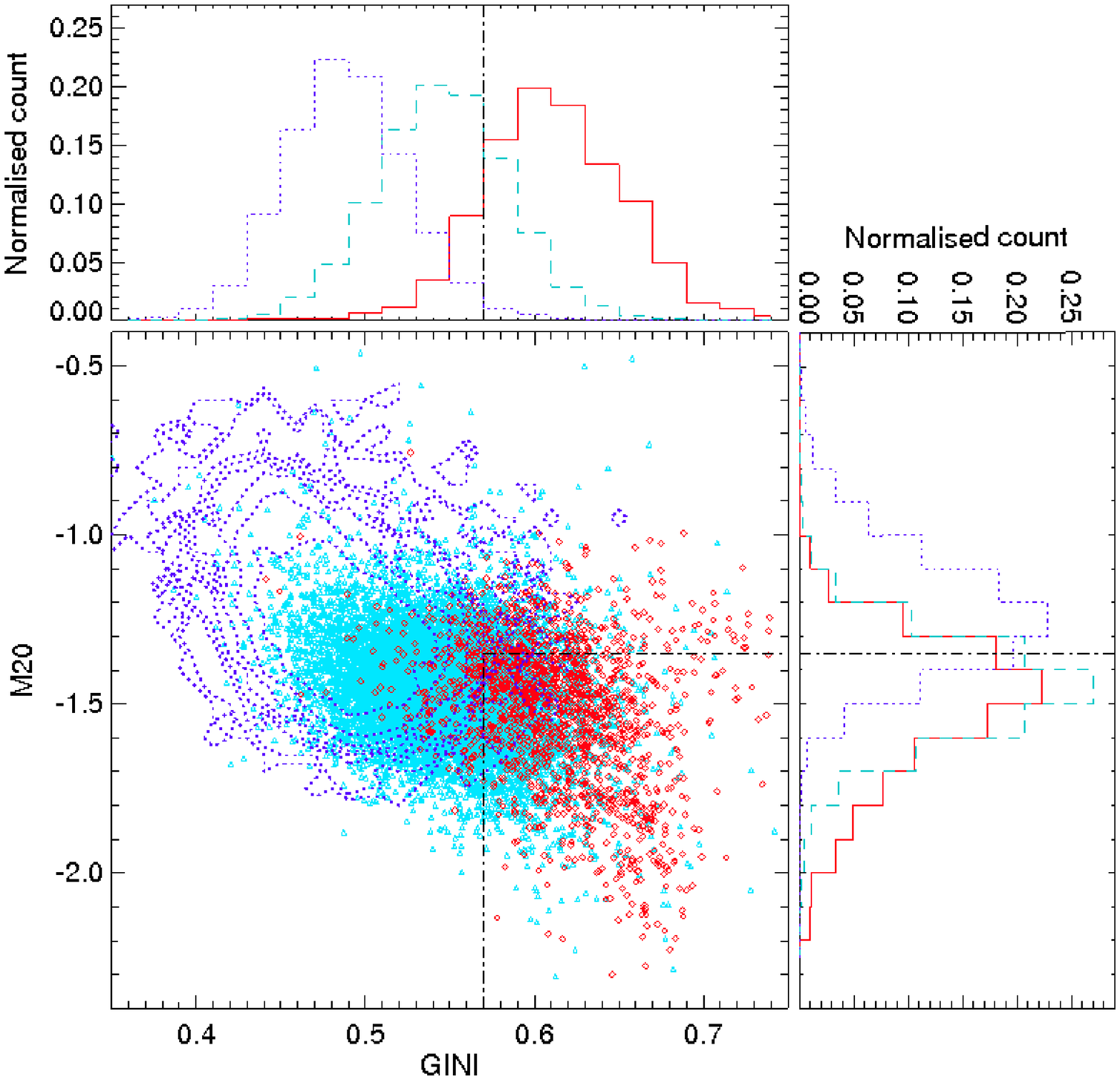}
\end{minipage}
\begin{minipage}[c]{.49\textwidth}
\includegraphics[width=0.85\textwidth,angle=0]{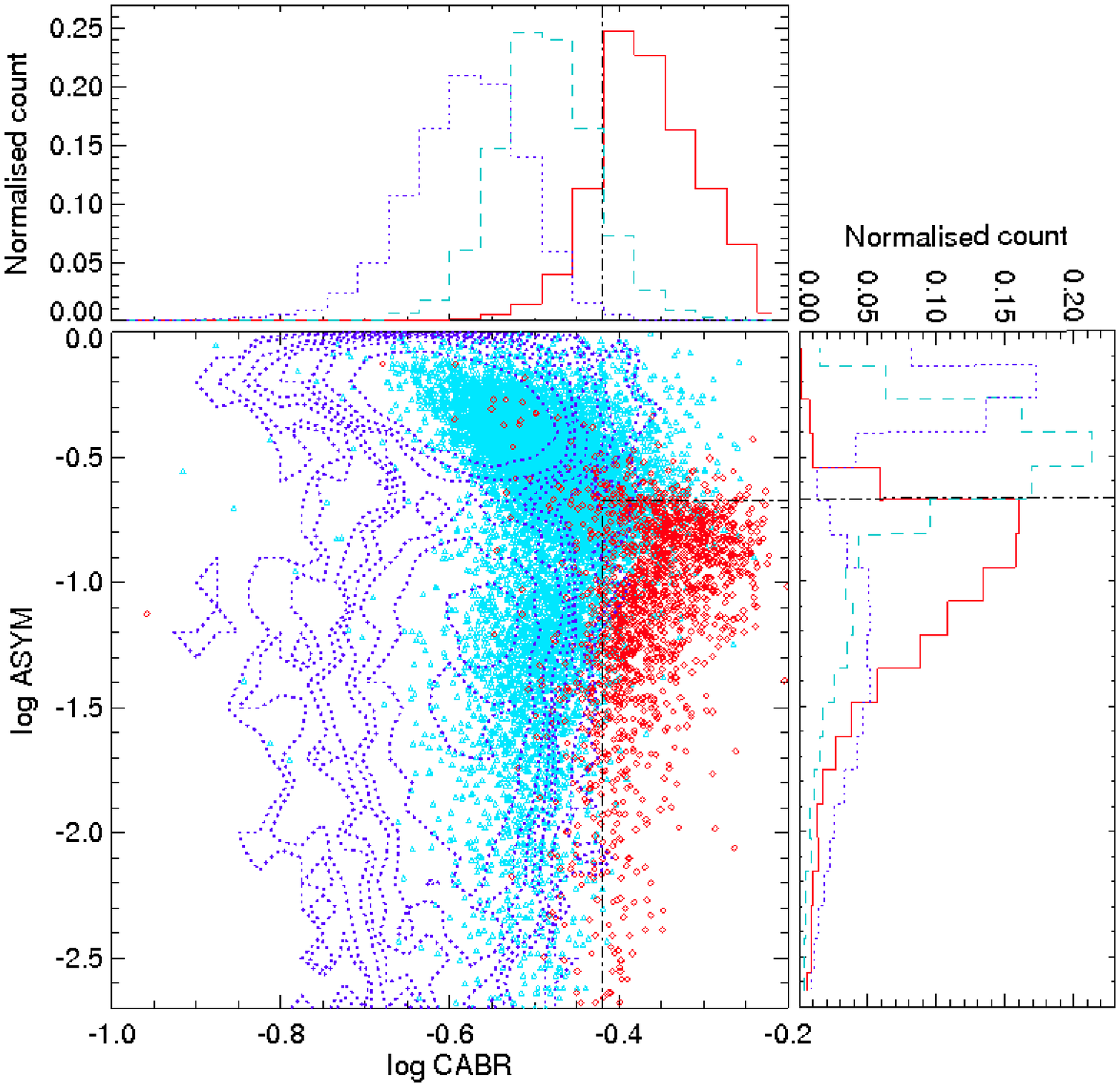}
\end{minipage}
\begin{minipage}[c]{.49\textwidth}
\includegraphics[width=0.85\textwidth,angle=0]{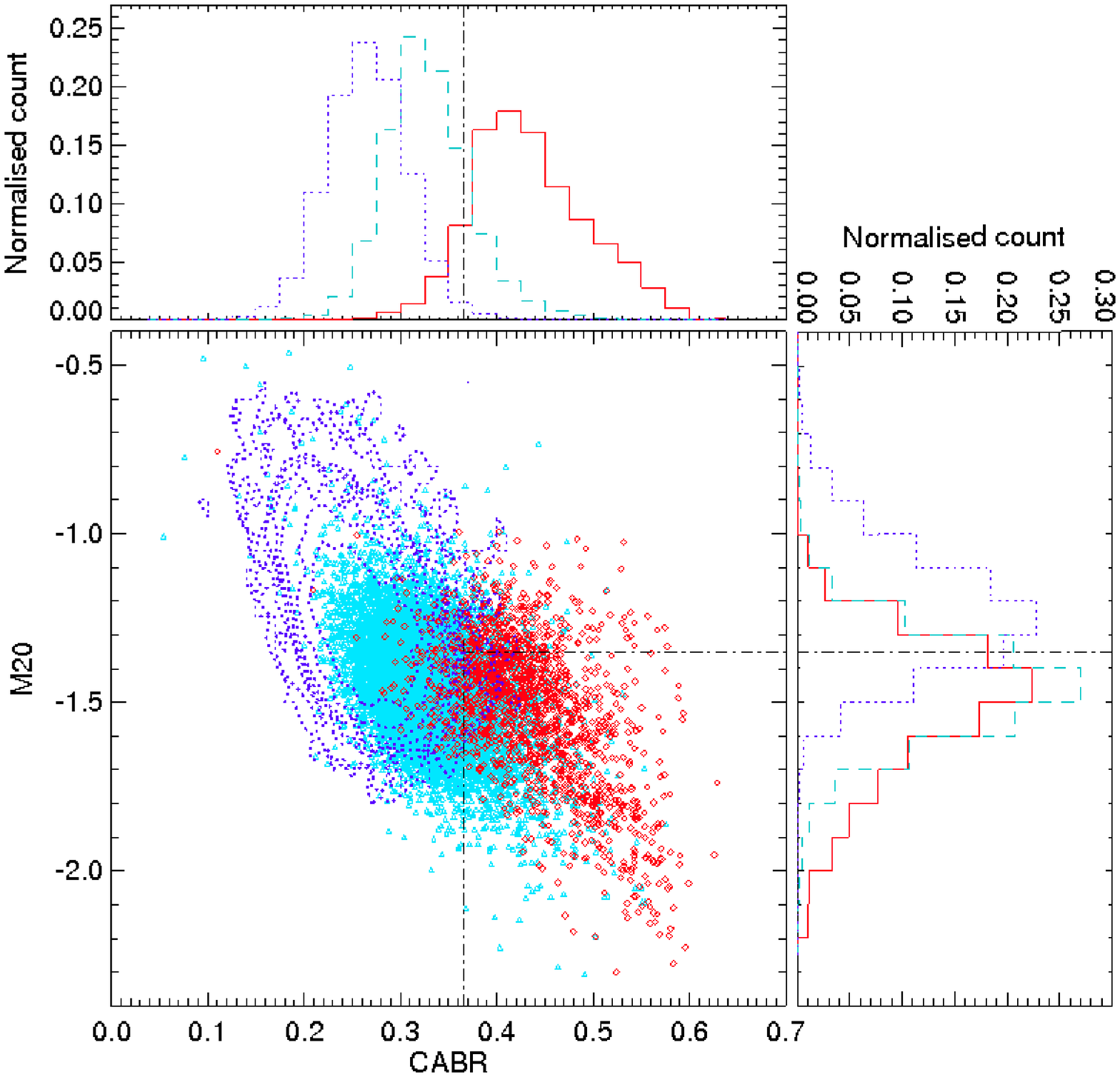}
\end{minipage}
\begin{minipage}[c]{.49\textwidth}
\includegraphics[width=0.85\textwidth,angle=0]{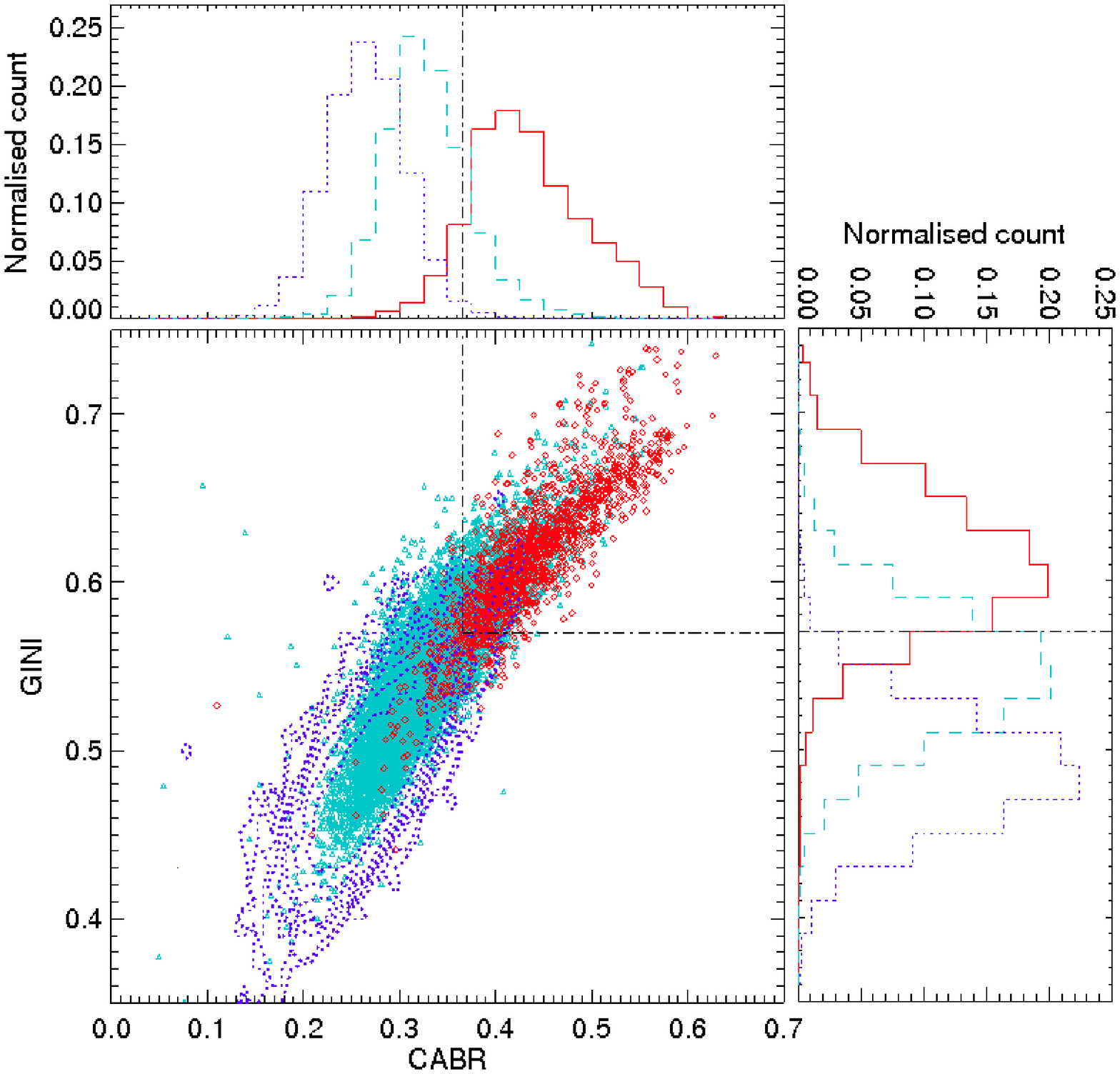}
\end{minipage}
\begin{minipage}[c]{.49\textwidth}
\includegraphics[width=0.85\textwidth,angle=0]{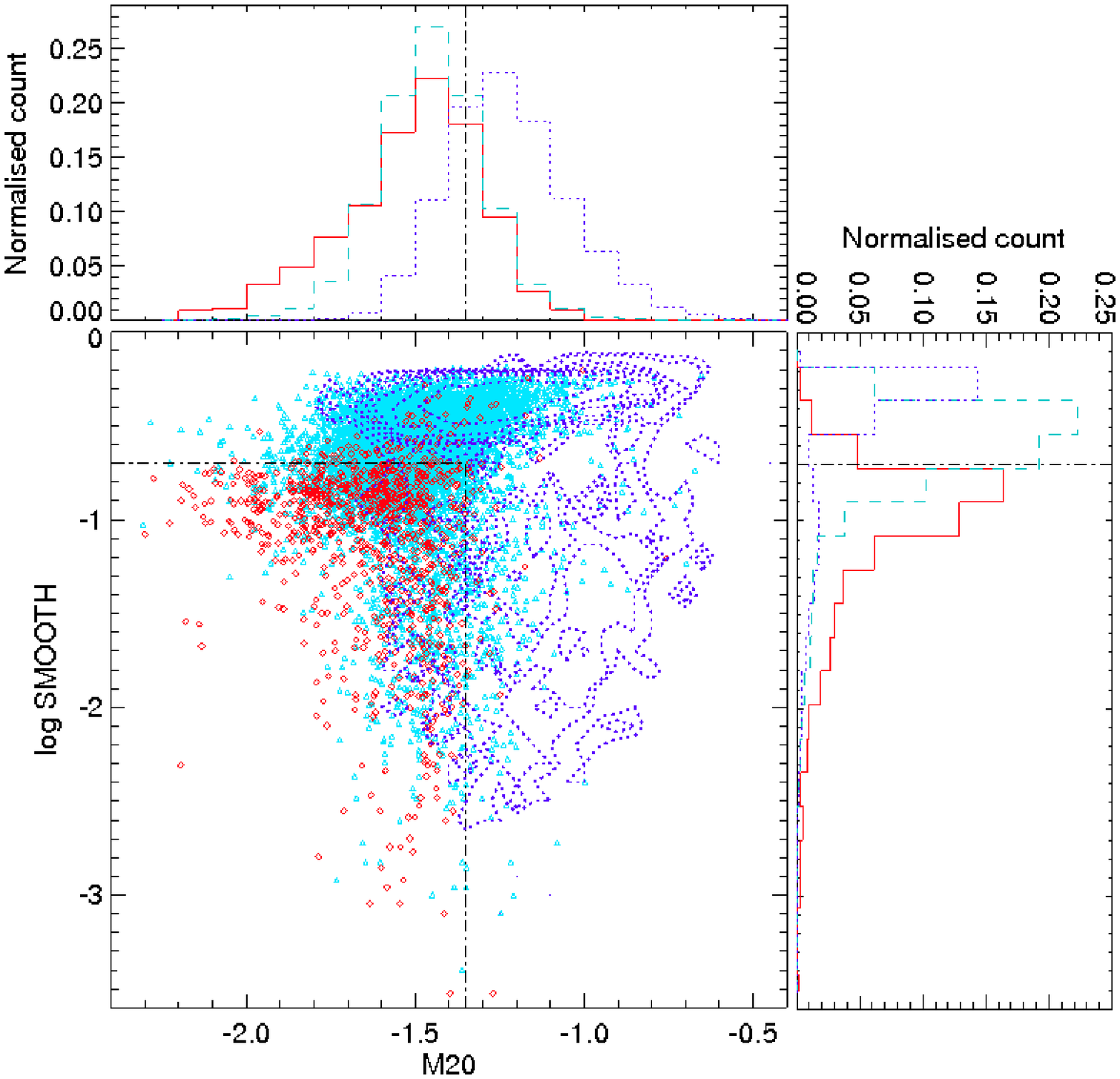}
\end{minipage}
\caption[ ]{\textit{Central plots:} morphological diagrams representing the relation between the logarithms of smoothness and Gini coefficient \textit{(top, left)}, M20 moment of light and Gini \textit{(top, right)}, logarithms of asymmetry and Abraham concentration indices \textit{(middle, left)}, M20 moment of light and Abraham concentration index \textit{(middle, right)}, Gini and Abraham concentration index \textit{(bottom, left)}, and finally, logarithm of smoothness vs. M20 moment of light \textit{(bottom, right)}. In all plots, of p10 sources classified down to F613W\,$\le$\,22.0, ET galaxies are marked with red diamonds, and LT with blue triangles (darker and brighter symbols, respectively, when printed in black and white). Boxes marked with dash-dot-dashed black lines define the locus of $\sim$\,80\% and $\sim$\,20\% of galaxies from ET and LT samples, respectively. \textit{Top and right histograms:} normalised distributions of corresponding parameters represented on the central plots of ET (red solid lines) and LT (blue dashed lines) galaxies. Dash-dot-dashed black lines showed on the histograms are values that separate the majority of sources in two classes. Green dotted contours and histograms represent the distributions of p10 LT galaxies with magnitudes 22.0\,$<$\,F613W\,$\le$\,23.0. 
\label{fig_morph_diagrams}}
\end{figure*}

\indent The $p_{10}$ thresholds and hence the morphological classification were calibrated in the ALH-4 field, and extrapolated to the other fields. We checked in Fig.~\ref{fig_morph_diagrams} that the classified galaxies on the full survey populate the expected regions in the morphological planes. The central panels, from left to right and from top to bottom, show relations between the smoothness and gini, M20 moment of light and gini, asymmetry and Abraham concentration index, M20 moment of light and Abraham concentration index, gini and Abraham concentration index, and finally, smoothness and M20. In all plots, we present F613W\,$\le$\,22.0 p10 sample, where ET galaxies are marked with red diamonds, and LT with blue triangles. Top and right panels of each diagram, represent normalised distributions of corresponding parameters for ET (red solid lines) and LT (blue dashed lines) galaxies. In addition, violet contours in the central diagrams and dotted violet histograms represent the distribution of 22.0\,$<$\,F613W\,$\le$\,23.0 p10 galaxies classified as LT (see Section~\ref{sec_final_class}). In all central plots and histograms there is a clear separation between the two morphological types. It is well known that, on average, ET galaxies present higher central light concentrations (CABR, GINI, and CCON), but lower asymmetries and small-scale structures (SMOOTH, ASYM, M20) than LTs. These expected trends are observed in the full sample which confirms that our $p_{10}$ morphological classification is reliable in all ALH fields. Areas marked with black dash-dot-dashed lines on the central plots define the regions populated by $\sim$\,80\% of ET and $\sim$20\% of LT galaxies in our sample, down to magnitudes 22.0 in the F613W band. For each morphological parameter, we report in Table~\ref{tab_morph_param_limits} the criteria obtained from normalized distributions to isolate most of the ET and LT galaxies, and also indicate for each region the expected population of both classes.

\begin{table*}
\begin{center}
\caption{Distribution of morphological parameters of ET and LT galaxies. 
\label{tab_morph_param_limits}}
\begin{tabular}{c | c | c | c | c | c}
\hline
\textbf{Parameter}&\textbf{ET criteria}&\textbf{ET pop.}&\textbf{LT criteria}&\textbf{LT1$^*$ pop.}&\textbf{LT2$^{**}$ pop.}\\ 
\hline
GINI&$>$\,0.57&85\%&$<$\,0.57&74\%&98\%\\
log SMOOTH&$<$\,-0.70&88\%&$>$\,-0.70&46\%&22\%\\
M20&$<$\,-1.35&78\%&$>$\,-1.35&25\%&75\%\\
CABR&$>$\,0.36&92\%&$<$\,0.36&79\%&99\%\\
log ASYM&$<$\,-0.67&86\%&$>$\,-0.67&62\%&44\%\\
CCON&$>$\,2.40&75\%&$<$\,2.40&84\%&95\%\\ 
\hline
\end{tabular}
\end{center}
\begin{flushleft}
{* F613W\,$\le$\,22.0\\
** 22.0\,$<$\,F613W\,$\le$\,23.0}
\end{flushleft}
\end{table*}

\subsection[]{Absolute magnitudes, stellar masses, and colours}
\label{subsec_stm_mag_z}

\indent We measured the K-corrections by means of the IDL routine \texttt{KCORRECT} \citep{blanton07}. We implemented all 23 ALH filters and their response files into the code, fitting our 23 point SED with about 500 available spectral templates \citep[see][for more information about the templates and SED fitting]{blanton07}. 91\%, 98\%, and 93\% of sources have the difference between the original magnitudes and those recovered from the final KCORRECT fits below 0.3, 0.2, and 0.2 in three observed bands, respectively (or 98\%, 100\%, and 100\% if the difference is $<$\,0.5), showing a good quality of our K-corrections. \\
\indent We then obtained the rest-frame magnitudes, absolute magnitudes, rest-frame colours, and luminosities of all sources presented in the catalogue we described above. Moreover, from the \texttt{KCORRECT} SED fits we obtained the stellar masses and star-formation histories, using the \cite{chabrier03} stellar initial mass function. We do not have any mid-infrared (MIR) nor far-infrared (FIR) data as an input information for SED fitting, the reddest band we have is $K$. This means that, following the well known mathematical relation between the observed and rest-frame wavelengths (1\,+z\,=\,$\lambda$\,$_{observed}$/$\lambda$\,$_{emited}$), we can count with our K-corrections and measured absolute magnitudes and stellar masses down to redshifts $\sim$\,1.3 in all 20 optical bands, and down to $\sim$\,0.8 and $\sim$\,0.4 in the $J$ and $H$ NIR bands, respectively. We compared obtained stellar masses with those measured with BPZ code. The typical scatter between masses is 0.22\,dex for all $p_{10}$ classified galaxies, or 0.17\,dex for ETs and 0.21\,dex for LTs. Stellar masses in all ALH survey measured with BPZ will be released in a companion paper \citep{molino13}. 

\begin{figure*}
\centering
\begin{minipage}[c]{.49\textwidth}
\includegraphics[width=8.4cm,angle=0]{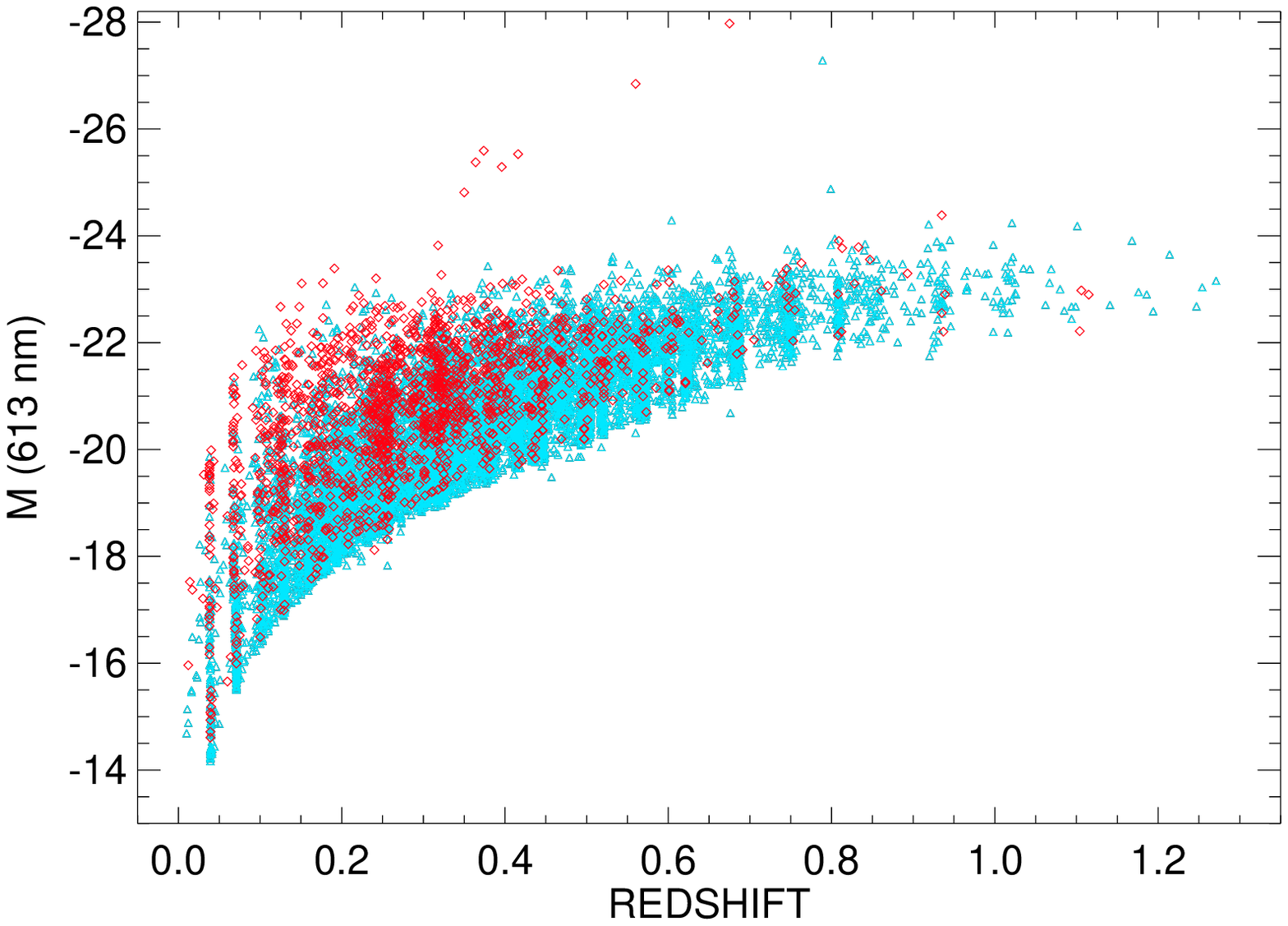}
\end{minipage}
\begin{minipage}[c]{.49\textwidth}
\includegraphics[width=8.35cm,angle=0]{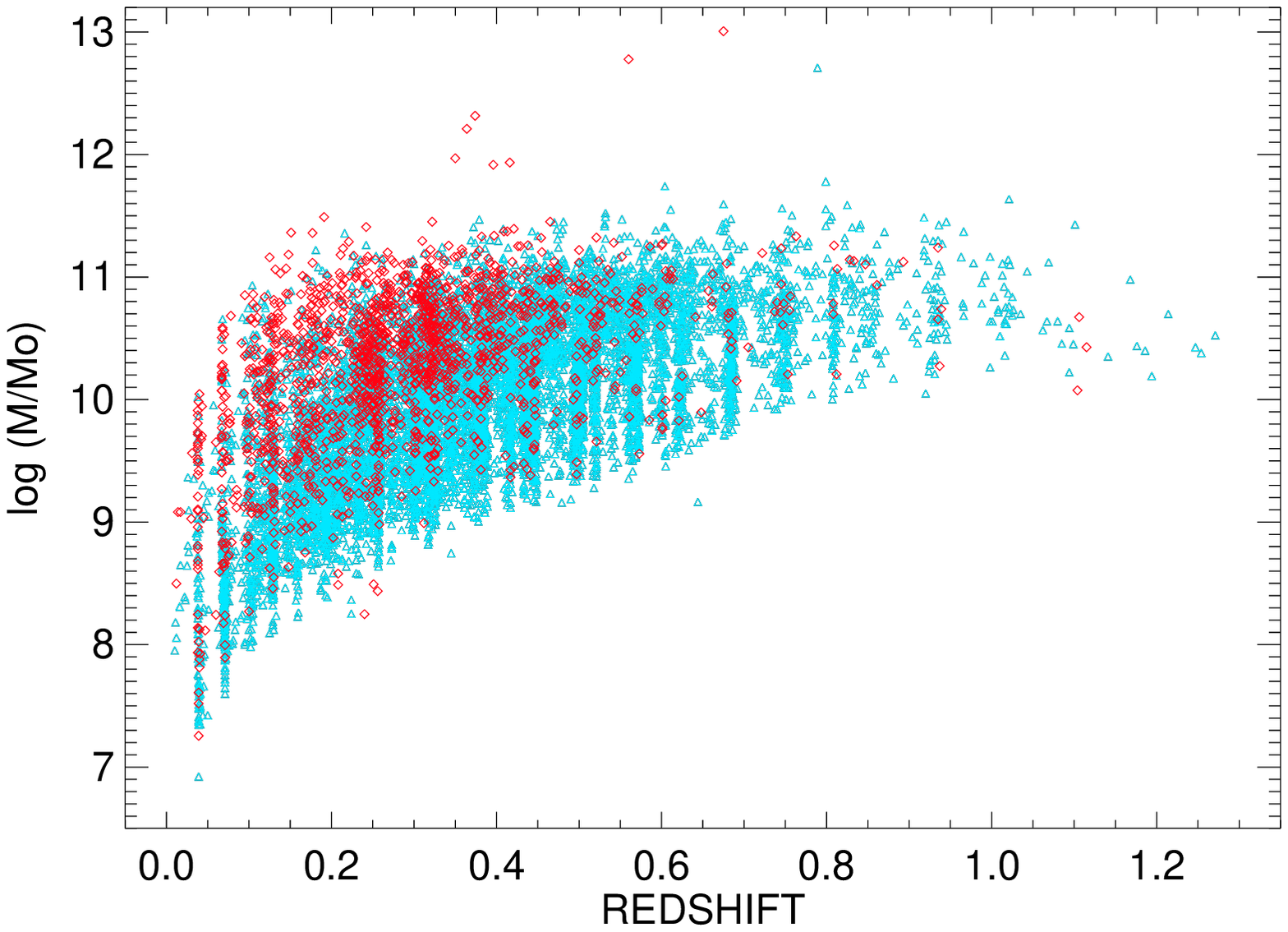}
\end{minipage}
\caption[ ]{Relation between the redshift and F613W absolute magnitude \textit{(left)}, and logarithm of stellar mass in solar mass units \textit{(right)} for ET (red diamonds) and LT (blue triangles) galaxies down to magnitudes F613W\,$\le$\,22.0.
\label{fig_absmag_stmass_z_morph}}
\end{figure*}

\indent Figure~\ref{fig_absmag_stmass_z_morph} shows the F613W band absolute magnitude\,-\,redshift (left panel) and stellar mass\,-\,redshift (right panel) relations of ET and LT galaxies (with F613W\,$\le$\,22.0). Notice the lack of ETs at higher redshifts which is a consequence of the selection effects discussed in section~\ref{subsec_sel_effects}. Compared to COSMOS/HST data, we trace the same population of late-type galaxies (see Fig. 1 and 2 in Tasca et al. 2009, and Fig. 2 in Pannella et al. 2009). However, for ET galaxies we are incomplete including lower redshifts. We can not distinguish ET sources from LTs fainter than -18.0 or -20.0 in the F613W band at redshifts 0.2 or 0.5, respectively. 

\begin{figure*}
\centering
\begin{minipage}[c]{.49\textwidth}
\includegraphics[width=8.4cm,angle=0]{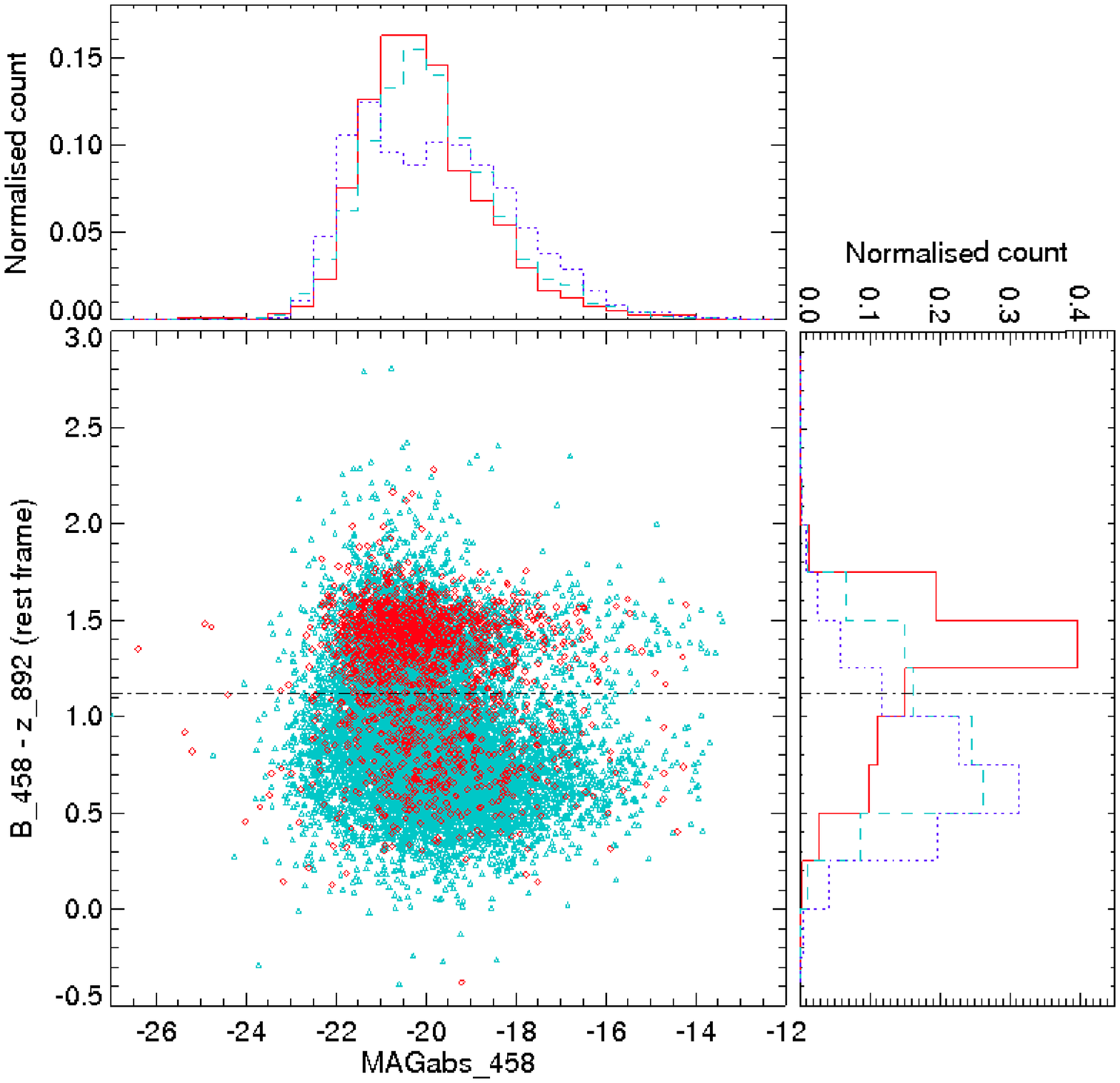}
\end{minipage}
\begin{minipage}[c]{.49\textwidth}
\includegraphics[width=8.35cm,angle=0]{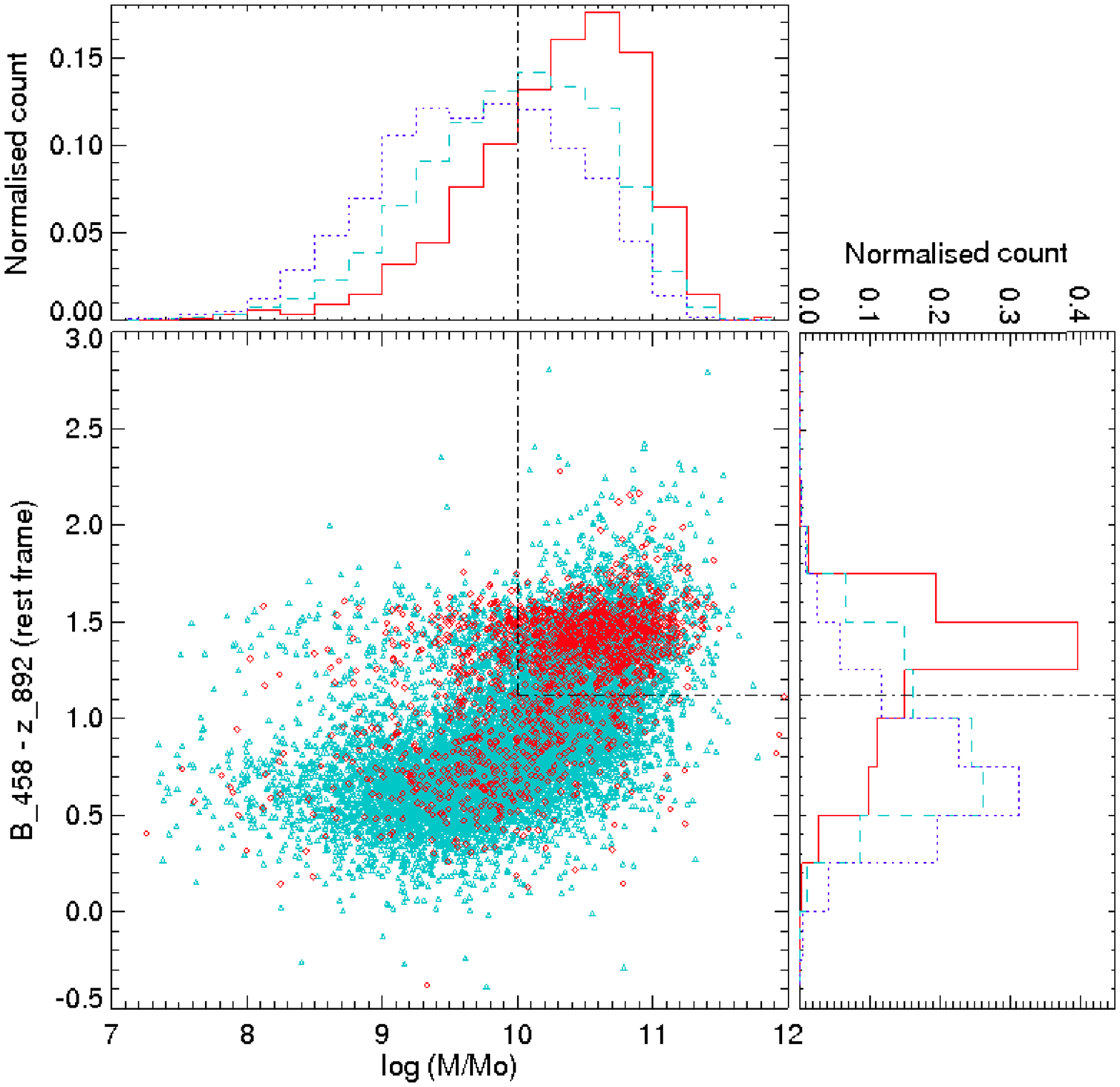}
\end{minipage}
\caption[ ]{\textit{(Left:)} Relation between the rest-frame $B$\,-\,$z$ colour and absolute magnitude in F458W band (central diagram) of ET (red diamonds) and LT (blue triangles) galaxies down to magnitudes F613W\,$\le$\,22.0.. To estimate the colour we used the information from the F458W and F892W ALH bands. Histograms present the normalised distributions of compared parameters: absolute magnitude (above the central plot) and colour (to the right of the central diagram) for ET (red solid lines) and LT (blue dashed lines) sources. Dotted violet histograms show the distributions of LT galaxies with magnitudes 22.0\,$<$\,F613W\,$\le$\,23.0. \textit{(Right:)} Relation between the rest-frame $B$\,-\,$z$ colour and stellar mass (central diagram). Histograms show the normalised distributions of analysed parameters, as in previous case. All symbols and lines have the same significance as in left diagram.
\label{fig_cmr_csmr_morph}}
\end{figure*}

\indent In Fig.~\ref{fig_cmr_csmr_morph} we tested the distributions of our p10 F613W\,$\le$\,22.0 ET (red diamonds and solid-lines) and LT (blue triangles and dashed-lines) selected galaxies on standard colour-magnitude and colour-stellar mass diagrams. We measured the rest-frame colour between the F458W and F892W ALH bands, which corresponds to approximately central wavelengths of standard Johnson $B$ and SDSS $z$ bands. This colour is compared with the absolute magnitude in the F458W band (left panel) and stellar mass (right panel). We also present, for both morphological types, the histograms with normalised distributions of corresponding parameters, including also the p10 LT galaxies selected in the magnitude range 22.0\,$<$\,F613W\,$\le$\,23.0. In all diagrams, we can see a clear bimodal distribution between the 'red sequence' and 'blue cloud' galaxies, widely discussed in the literature \citep[e.g.,][are only some of them]{strateva01,hogg03,bell03,melbourne07,cassata07}, where the majority of ET and LT galaxies are located, respectively. This bimodal distribution of galaxies on both colour-magnitude and colour-stellar mass diagrams is sometimes used to distinguish between the early- and late-type populations of galaxies \citep[e.g.][]{bell03,faber07,franzetti07}. However, it was showed that a fraction of spirals and irregulars with high extinctions or quenched stellar formations can reside in the red sequence, or earlier-types with bursts of star formation in the blue cloud \citep[e.g.,][]{williams08,bell08,torre11,oteo13a,oteo13b}. Black horisontal line in Fig.~\ref{fig_cmr_csmr_morph} shows the regions where 70\% of our F613W\,$\le$\,22.0 ET and LT p10 galaxies lie in the red sequence (B458\,-\,z892\,$>$\,1.12) and in the blue cloud (B458\,-\,z892\,$<$\,1.12; 85\% of LT galaxies selected with magnitudes 22.0\,$<$\,F613W\,$\le$\,23.0), respectively. Through the colour-stellar mass relation in Fig.~\ref{fig_cmr_csmr_morph} (left panel), we defined the locus where $\sim$\,60\% of our ET galaxies are located, having B458\,-\,z892 colour $>$\,1.12 and stellar masses log\,M/M$_o$\,$>$\,10.0. On the other side, $\sim$\,25\% and $\sim$\,10\% of F613W\,$\le$\,22.0 and 22.0\,$<$\,F613W\,$\le$\,23.0 LT galaxies reside in this region.

\section[]{Summary and conclusions}
\label{sec_summary}

\indent We presented the morphological classification of $>$\,40,000 galaxies in the ALHAMBRA survey (in seven fields), classifying all galaxies in two broad morphological classes: early- and late-types. With this paper we release the low-contamination catalogue of 22,051 galaxies classified with the contamination lower than 10\%. We classified 1,640 and 10,322 early- (down to redshifts $\sim$\,0.5) and late-type (down to redshifts $\sim$\,1.0) galaxies, respectively, with magnitudes F613W\,$\le$\,22.0. In addition, for magnitude range 22.0\,$<$\,F613W\,$\le$\,23.0 we classified other 10,089 late-type galaxies with redshifts $\le$\,1.3. ALHAMBRA is a photometric survey, having for all detections observations in 23 optical and NIR bands. With a large number of detected sources ($>$\,670,000) down to photometric completeness of SDSS $r$\,$\sim$\,25.0, large covered area, and precisely measured photometric redshifts (obtained through SED fittings of 23 point spectra), ALHAMBRA is an ideal survey for tracing the cosmic variance and cosmic evolution. However, morphological properties and classification of galaxies are crucial for any kind of galaxy formation and evolution studies. Low-contamination, high-populated morphological catalogue presented in this paper, together with the precise measured photometric redshifts and photometric properties of the ALH survey, present an important addition to other datasets for studying the morphological properties of extragalactic sources and their evolution (taking into account the variety of covered ALH fields). Some of these studies include: the evolution of ETs and LTs down to redshifts $\sim$\,0.5 and $\sim$\,1.3, respectively, their star-formation histories, morphological properties and evolution of active galaxies and their comparison with non-active galaxies, as well as statistical comparisons between the morphological and SED fitting classifications. Using the ground based data, Galaxy Zoo survey classified morphologicaly $\sim$\,900,000 galaxies down to redshifts $\sim$\,0.25 \citep{lintott11}. On the other side, COSMOS survey provides morphological classification of $>$\,200,000 galaxies \citep{tasca09} at high-redshifts. With this work we provide the astronomical society with the additional morphological information of $>$\,22,000 high-redshift galaxies, but observed in seven different fields which may have an important constraints on cosmic variance and galaxy evolution studies.\\
\indent To select the sample for morphological classification, we first separated galaxies from point-like sources, then we selected sources with good photo-z measurements (observed in all filters and with BPZ\_ODDS parameter $>$\,0.2), and finally, we selected sources photometrically, considering only detections with magnitudes $\le$\,23.0 in the F613W band and photometric errors $<$\,0.5 (to make sure to deal with reliable extended vs. point-like source selections, photometric, and photo-z measurements). The final selected sample has 43,665 sources. \\
\indent We used the galSVM code in our classification, testing the morphology through a sample of 3,000 visually classified local galaxies. We redshifted these galaxies and scaled them in luminosity, to reproduce the redshift and magnitude distributions of our ALH sources. Morphological classification was carried out in the F613W band, the most efficient one of all 20 optical bands, and for six magnitude cuts, down to 23.0. For each ALH galaxy we measured 7 morphological parameters and the averaged $p_{10}^E$ probability that the galaxy is ET (probability that the galaxy is LT is then $p_{10}^L$\,=\,1\,-\,$p_{10}^E$), obtained through 15 Monte-Carlo simulations.\\
\indent Our classification is calibrated against the COSMOS field, using the HST/ACS images. We used this calibration to determine the probability cuts (in each of six magnitude cuts) to select ET and LT galaxies with the contamination lower than 10\%. With the obtained probability cuts, we can recover $\sim$\,70\% of ET galaxies down to magnitudes 20.0, 30\,-\,40\% down to 21.5, and 20\,-\,30\% down to 22.0 in the F613W band. On the other side, for LT galaxies, we recover $\sim$\,70\% down to magnitudes 22.0, $\sim$\,60\,-\,70\%, down to 22.5, and $\sim$\,30\% down to 23.0. We tested our classification in whole ALH survey through different morphological diagrams and general ET and LT relations (e.g., colour-magnitude and colour-stellar mass diagrams), obtaining the expected distributions in all of analysed relations.\\    
\indent The complete, low-contamination ($<$\,10\%) catalogue of 22,051 galaxies provides all measured morphological parameters, averaged probabilities, morphological types, magnitudes, physical sizes, and redshifts. The catalogue is available in the electronic version of this paper and through the ALHAMBRA webpage http://alhambrasurvey.com/, while the description of columns and the small example of five sources are presented in the Appendix.

\section*{Acknowledgments}

We thank the referee Chris Lintott for constructive comments which improved the paper significantly. This research was supported by the Junta de Andaluc\'ia through projects PO8-TIC-03531 and TIC114, the Spanish Ministry of Economy and Competitiveness (MINECO) through projects AYA2006-14046, AYA2010-15169, AYA2010-22111-C03-02, AYA2011-29517-C03-01, and the Generalitat Valenciana through project GV/Prometeo 2009/064. MP acknowledge financial support from JAE-Doc program of the Spanish National Research Council (CSIC), co-funded by the European Social Fund. Based on observations collected at the Centro Astron\'omico
Hispano Alem\'an (CAHA) at Calar Alto, operated jointly by the Max-Planck Institut fur Astronomie and the Instituto de Astrof\'isica de Andaluc\'ia (CSIC). The CEFCA is funded by the Fondo de Inversiones de Teruel, supported by both the Government of Spain (50\%) and the regional Government of Arag\'on (50\%). In this work, we made use of Virtual Observatory Tool for OPerations on Catalogues And Tables (TOPCAT). We thank the developers of Ubuntu, Python, Numpy, Scipy, and Matplotlib for making their work public and available to all scientific community.

\appendix

\section[]{ALHAMBRA morphological catalogue}
\label{appendixA_cat}

\indent In this section, we describe the high-quality, two-class morphological catalogue in the ALHAMBRA survey, with the total contamination lower than 10\%. Catalogue contains morphological, photometric, size, and photometric redshift information of 22,051 galaxies. Of those, 1,640 and 20,411 were classified as early- and late-types, down to magnitudes 22.0 and 23.0 and photometric redshifts of $\sim$\,0.5 and $\sim$\,1.3, respectively\footnote{In case you need less-strict catalogue, with the contamination above 10\%, or the total one with 43,665 sources down to magnitudes 23.0, as well as any additional information presented in the paper but not available in the published catalogue, please contact us at mpovic@iaa.es}. Table~\ref{tab_general_morph} shows an example of the format and content of the catalogue for five sources. The catalogue is available in the electronic addition of this paper or through the ALHAMBRA website http://alhambrasurvey.com/. The column entries are as follows:

\begin{table*}
\begin{center}
\caption{Morphological catalogue of 22,051 galaxies classified  in the ALHAMBRA survey with the contamination $<$\,10\%
\label{tab_general_morph}}
\begin{tabular}{c c c c c c c}
\noalign{\smallskip}
\hline
\textbf{ID}&\textbf{FIELD\_P\_CCD}&\textbf{ID\_phot}&\textbf{ID\_zphot}&\textbf{RA (degrees)}&\textbf{DEC (degrees)}&\textbf{FLAGS}\\
\textbf{m$_{458}$}&\textbf{errm$_{458}$}&\textbf{m$_{613}$}&\textbf{errm$_{613}$}&\textbf{m$_{892}$}&\textbf{errm$_{892}$}&\textbf{logR50 (kpc)}\\
\textbf{logR90 (kpc)}&\textbf{MUMEAN}&\textbf{ELLIPTICITY}&\textbf{ASYM}&\textbf{CABR}&\textbf{GINI}&\textbf{SMOOTH}\\
\textbf{M20}&\textbf{CCON}&\textbf{$p_{10}^E$\_AVG}&\textbf{err$p_{10}^E$\_AVG}&\textbf{CLASS}&\textbf{REDSHIFT}&\\
\hline
1 &ALH2\_p1\_ccd1& 8895&81421100234&37.43648& 1.264531&0\\
23.0&0.098&22.04&0.022&21.067& 0.084&0.723\\
0.985&24.5309&0.2143&0.2686&0.3075&0.532&0.0\\
-1.2281&2.1533&0.3034&0.0417&LT&0.666&\\
&&&&&\\	
2 &ALH2\_p1\_ccd1& 9202&81421100256&37.523216&1.263539&2\\
23.565&0.231&22.411& 0.04& 21.052& 0.125&0.971\\
1.356&25.5367&0.4043&0.8212&0.2458&0.4558&0.633\\
-0.9595&2.3292&0.4649&0.1535&LT&0.598&\\
&&&&&\\	
3 &ALH2\_p1\_ccd1& 9353&81421100262&37.532192&1.263552&2\\
23.033&0.122&22.323& 0.032&21.524& 0.16&0.879\\
1.125&25.3686&0.1158&0.0211&0.2054&0.4731&0.0\\
-0.9625&1.8086&0.3041&0.0437&LT&0.681&\\
&&&&&\\	
4 &ALH2\_p1\_ccd1& 9453&81421100274&37.582558&1.262976&0\\
22.042&0.069&20.741& 0.012&19.914& 0.056&0.749\\
1.094&24.1509&0.1518&0.1118&0.3918&0.5719&0.1679\\
-1.4883&2.6371&0.8955&0.0948&ET&0.421& \\
&&&&&\\	
5 &ALH2\_p1\_ccd1& 8568&81421100275&37.398445&1.263138&2\\
22.036&0.051&21.104& 0.012&20.657& 0.079&0.564\\
0.876&24.0459&0.4405&0.053&0.3176&0.5584&0.1136\\
-1.5274&2.2498&0.138&0.0389&LT&0.258&\\
\noalign{\smallskip}
\hline
\end{tabular}
\end{center}
\end{table*}

\begin{itemize}
\item Column 1 (ID): Identification number.

\item Column 2 (FIELD\_P\_CCD): ALHAMBRA field, pointing, and CCD.

\item Column 3 (ID\_phot): Identification number in the photometric catalogue \citep{husillos13}; equal to NUMBER parameter in the original catalogue.

\item Column 4 (ID\_zphot): Identification number in the photometric redshift catalogue \citep{molino13}; equal to ID parameter in the original catalogue.

\item Columns 5 and 6 (RA, DEC): Equinox J2000.0 right ascension and declination in degrees of the centroid.

\item Column 7 (FLAGS): SExtractor FLAG parameter \citep{bertin96} contained in the photometric catalogue.

\item Columns 8 and 9 (m$_{458}$, errm$_{458}$): Apparent magnitude in the F458W band and its error.

\item Columns 10 and 11 (m$_{613}$, errm$_{613}$): Apparent magnitude in the F613W band and its error.

\item Columns 12 and 13 (m$_{892}$, errm$_{892}$): Apparent magnitude in the F892W band and its error.



\item Column 14 (logR50): logarithm of radius at 50\% of flux in kpc.

\item Column 15 (logR90): logarithm of radius at 90\% of flux in kpc.

\item Column 16 (MUMEAN): Mean Surface Brightness measured by galSVM.

\item Column 17 (ELLIPTICITY): ellipticity parameter measured by SExtractor.

\item Column 18 (ASYM): Asymmetry index measured by galSVM, defined as in \cite{conselice00}.

\item Column 19 (CABR): Abraham concentration index, measured by galSVM and defined as in \cite{abraham96}.

\item Column 20 (Gini): Gini coefficient measured by galSVM and defined as in \cite{abraham03}.

\item Column 21 (SMOOTH): Smoothness of the source, measured by galSVM and defined as in \cite{conselice00}. 

\item Column 22 (M20): Moment of light at 20\%, measured by galSVM and defined as in \cite{lotz04}. 

\item Column 23 (CCON): Conselice-Bershady concentration index, measured by galSVM and defined as in \cite{bershady00}. 

\item Columns 24 and 25 ($p_{10}^E$\_AVG and err$p_{10}^E$\_AVG): Averaged probability that the galaxy is ET, and its error. The probability is measured from other 15 probabilities (see Sec.~\ref{subsec_galsvm_config_classifiaction}). It takes values from 0 to 1, where small values indicate that the galaxy is LT. We measure $p_{10}^L$ probability as $p_{10}^L$\,=\,1\,-\,$p_{10}^E$.

\item Column 26 (CLASS): Final morphological class, after applying the probability cuts described in Sec.~\ref{sec_class_reliability}.

\item Column 27 (REDSHIFT): Photometric redshift \citep{molino13}


\end{itemize}

\label{lastpage}

\end{document}